\documentclass[article,aps,prd,12pt,notitlepage]{revtex4-2}

\usepackage{bm}
\usepackage{amssymb}
\usepackage{graphicx,xcolor}
\usepackage{amsmath,graphicx}
\usepackage{csquotes}
\usepackage{amsmath,amssymb,stmaryrd}
\usepackage{dsfont}
\usepackage{braket}
\begin{document}
\def\la{{\langle}}
\def\e{\enquote}
\def\u{\hat u}
\def\U{\hat U}
\def\A{\hat A}
\def\AA{\mathcal{ A}}
\def\B{\hat B}
\def\C{\hat C}
\def\k{^{KG}}
\def\f{\phi}
\def\Om{\Omega}
\def\up{\uparrow}
\def\do{\downarrow}
\def\ep{\epsilon}
\def\fb{\overline F}
\def\wb{\overline W}
\def\nl{\newline}
\def\h{\hat h}
\def\H{\hat H}
\def\Psii{\hat \Psi}
\def\pt{\partial_t}
\def\lm{\lambda}
\def\lmu{\underline\lambda}
\def\q{\quad}
\def\t{\tau}
\def\p{\hat\pi}
\def\ps{\hat\Psi}
\def\r{\color{red}}
\def\g{\color{green}}
\def\b{\color{blue}}
\def\n{\\ \nonumber}
\def\ra{{\rangle}}
\def\Ep{{\mathcal{E}}}
\def\Ep{{\mathcal{E}}}
\def\E{{\epsilon}}
\def\a{{\hat a}}
\def\b{{\hat b}}
\def\c{{\hat C}}
\def\uu{{\hat u}}
\def\al{{\alpha}}
\def\omm{\omega_F}
\def\ommm{\omega_B}
\def\be{{\beta}}
\def\R{\text {Re}}
\def\I{\text {Im}}
\def\vc{\text {vac}}
\def\rx{]_\xi}
\def\+{\dagger}
\def\ph{\varphi}
\def\om{\omega}
\def\pp{\phi^+_k}
\def\nm{\phi^-_k}
\def\rak{\rangle_{KG}}
\def\K{{KG}}
\def\rhoo{\hat \rho}
\def\qq{\q\q\q\q\q\q\q\q\q\q\q\q\q\q\q\q\q\q\q\q}
\def\1{\mathds {1}}
\def\V{|V|}
\def\({[}
\def\vac{\text{vac}}
\def\Vac{\text{Vac}}
\title{Quantum statistical effects in  one-particle densities:
scattering and pair production}
 \date\today
%
%
\author{X. Guti\'errez de la Cal$^{a,d}$}
\author{M. Pons$^{b,d}$}
\author{A. Matzkin$^{e}$}
\author{D. Sokolovski$^{a,c,d,*}$}
\affiliation{$^{a}$ Departamento de Qu\'imica-F\'isica, Universidad del Pa\' is Vasco, UPV/EHU,  48940, Leioa, Spain}
\affiliation{$^{b}$ Departamento de F\' isica Aplicada, Universidad del Pa\' is Vasco, UPV-EHU, Bilbao, Spain}
\affiliation{$^{c}$ IKERBASQUE, Basque Foundation for Science, E-48011 Bilbao, Spain}
\affiliation{$^{d}$ EHU Quantum Center, Universidad del Pa\' is Vasco, UPV/EHU, 48940 Leioa, Spain}
\affiliation{$^{e}$ Laboratoire de Physique Th\'eorique et Mod\'elisation, CNRS Unit\'e
8089, CY Cergy Paris Universit\'e, 95302 Cergy-Pontoise cedex, France}
\email{dgsokol15@gmail.com}
\begin{abstract}
\noindent
We study space-time resolved densities of particle-hole pairs
produced by an external time-dependent field acting on
non-interacting non-relativistic particles.
It is shown that, at least  in some cases, the densities
are not affected by Fermi-Dirac or Bose-Einstein statistics,
and are determined only by the initial state of the multi-particle system.
The second quantisation technique is extended to Dirac electrons and, with some modifications, to Klein-Gordon
bosons.
The difference in pair production in these two relativistic cases
is analysed in some detail.
 \date\today
\end{abstract}

%
%
\maketitle
%
%
\vspace{0.1cm}
\section{Introduction}
In \cite{Grobe1}  the authors used quantum field theory in order to evaluate spatially resolved densities
of electron and positron pairs produced  by a strong external field.  Pair production was found to be reduced
in the presence of an electronic wave packet propagating through the region where positrons are being produced.
In Refs.  \cite{{Grobe1},{Grobe2}} the method was extended to non-interacting Klein-Gordon bosons.
\newline
The questions posed in \cite{{Grobe1},{Grobe2},{Grobe3}} can equally be asked in a non-relativistic context, e.g.,
regarding the particle-hole production in semiconducting solids \cite{Semi}.
Non-relativistic analysis, relying on the second quantisation technique \cite{{Schw}, {FeynS}}
is particularly simple if the interaction between the particles can be neglected. This approximation is often made
in applications such as the theory of Bose-Einstein condensates (see, e.g., \cite{BJ}), or electronic point contacts (see e.g., \cite{Shm}).
If so, to predict the behaviour of a multi-particle system, subjected to an external field, and described by a symmetrised or antisymmetrised wave function,
it is sufficient to know the one-particle transition amplitudes. One-particle densities, which offer only a crude description of the system,
are often unaffected by the type of statistics obeyed by the particles. In this case one needs to consider full counting statistics in order to find a difference
between fermions and bosons.
\newline
The purpose of this paper is twofold. Firstly, we will study, in the non-relativistic case, spatially resolved particle and hole densities,
in order to establish whether these are affected by the Pauli exclusion principle. We will demonstrate that the answer must be no,
since the same densities would be produced if the fermions were to be replaced by bosons.
\newline
Secondly, we will extend the second quantisation procedure to Dirac electrons and, with a small modification,
to Klein-Gordon bosons. Unlike in the non-relativistic case, pair production rates for relativistic fermions and bosons
are known to be very different \cite{{Grobe2},{Matz}}, and we will study the origin of this discrepancy.
\newline
The rest of the paper is organised as follows. In Sect.II we will specify the condition under which it is meaningful
to consider the probabilities of finding, at a given location, a particle with positive, or negative energy.
In Sect.III we consider several distinguishable non-relativistic particles, each occupying one of the orthogonal states,
and evaluate the \e{particle} and \e{hole} densities once the external field is switched off.
In Sect.IV we replace the particles with non-relativistic fermions or bosons, and look for the difference in
the densities introduced in Sect.III.
In Sect.IV we use the hole theory of positrons to extend the treatment to Dirac electrons in one dimension.
In Sect.V we apply the second quantisation technique to Klein-Gordon bosons, and analyse both the similarities
and differences with the Dirac case.
In Sect.V we illustrate the difference between relativistic fermions and bosons by employing an
exactly solvable toy model.
Section VII contains our conclusions.
\section{A \e{double slit problem}}
\begin{figure}
\includegraphics[angle=0,width=10cm]{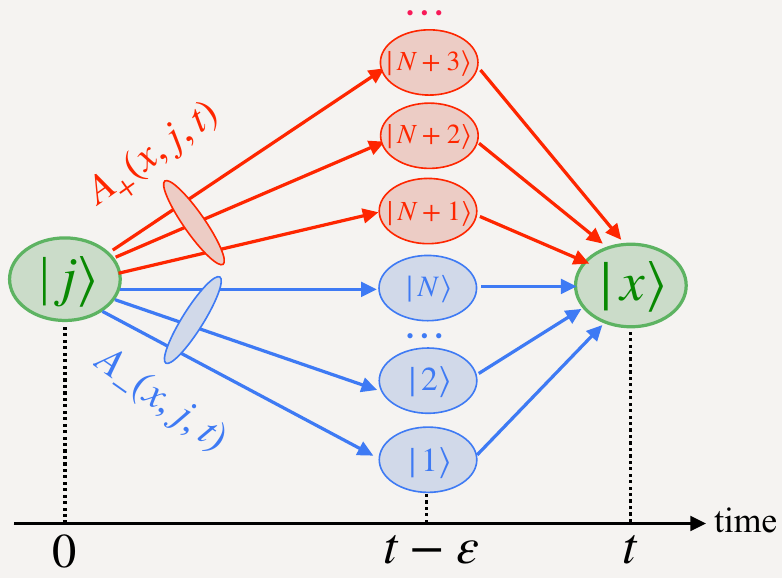}
\caption {The quantum system can reach a final state $|x\ra$ from an initial state $|j\ra$ by passing, immediately before, through one of the intermediate states, $|n\ra$, $E_n <0$, or $|p\ra$, $E_p >0$. The corresponding amplitude is $A(x\gets n(p)\gets j)= \la x|n(p)\ra\la n(p)|\u(t)|j\ra$. The amplitudes to be in $x$ and have a positive (negative) energy just before,  $A_{-(+)}(x,j,t)=\sum_{n(p)}=A(x\gets n(p)\gets j)$,are given in Eq.(\ref{3}).}
\label{FIG1}
\end{figure}

Consider a simple quantum system (e.g., a particle in an infinite rectangular well) with a Hamiltonian
$\h_0$,
and an energy spectrum $E_j$, $\h_0|j\ra =E_j |j\ra$ $j=1, 2,...$. Choosing the zero energy to lie between
$E_N$ and $E_{N+1}$ divides the spectrum into (n)egative  and (p)ositive parts,
$E_n=E_j$ for $j\le N$, and $E_p=E_j$ for $j>N$, as shown in Fig.1.
Using notations $|n\ra$, $n=1, ..., N$ and $|p\ra$, $p=N+1,N+2,...$
for the states corresponding to the negative and positive energies, respectively,
we will omit the limits of summations, and write $\equiv \sum_{n}=\sum_{n=1}^N$,
$\sum_{p}=\sum_{p=N+1}^\infty$, and $\sum_{j}=\sum_n+\sum_p$.
\newline
A  particle prepared in a state $|j\ra$ ($|p\ra$  or $|n\ra$) is subjected, after $t=0$, to an external field $\hat V(t)
=\sum_{jj'}|j\ra V_{jj'}(t)\la j'|$, which causes transitions between
the field-free states $|j\ra$. For  the Hamiltonian $\h(t)=\h_0 +\hat V(t)$ one has
\begin{eqnarray} \label{1}
\h(t)=\sum_{pp'}|p\ra h_{pp'}(t)\la p'|+\sum_{pn}|p\ra h_{pn}(t)\la n|
+\sum_{np}|n\ra h_{np}(t)\la p|+\sum_{nn'}|n\ra h_{nn'}(t)\la n'|,\q\q
\n
h_{jj'}(t) =h^*_{j'j}(t), \q j,j'=n,p.\qq\q\q\q\q
\end{eqnarray}
Once the field $V$ is switched off at a time $t$, what is the likelihood of a particle at $x$, to have, for example, positive energy?
\newline
The two properties are represented by projectors $\p_x=|x\ra\la x|$ and $\p_+ =\sum_{p}|p\ra\la p|$ which do not commute,
and cannot be established simultaneously. It is, however possible to measure one projector immediately after the other,
e.g., by employing a von Neumann pointer \cite{vN}.
With $\pi_+$ set to be measured first, one has a \e{double slit problem} shown in Fig.\ref{FIG1}.
The two amplitudes of the two pathways connecting $|j\ra$ at $t=0$ with $|x\ra$ at $t$ are given by
\begin{equation} \label{3}
\begin{aligned}
A_{+}(x,j,t)& = \la x|\pi_+ \u(t)| j\ra=\sum_p \la x|p\ra\la p| \uu(t) |j\ra= \la x |\p_+|j(t)\ra,\\
A_{-}(x,j,t) &=  \la x|\pi_- \u(t)| j\ra= \sum_n \la x|n\ra\la n| \uu(t) |j\ra = \la x |\p_-|j(t)\ra,
\end{aligned}
\end{equation}
with $|j(t)\ra\equiv \u(t)|j\ra$, $\u(t)= \exp[-i\int_0^t \h(t')dt']$ is the evolution operator (a time-orderd product is assumed where $\h(t')$ and $\h(t'')$ do not commute),
and
\begin{eqnarray} \label{31}
 \p_+\equiv \sum_{p}|p\ra\la p|, \q \p_-\equiv \sum_{n}|n\ra\la n|, \q  \p_++ \p_-=\hat 1.
 \end{eqnarray}
If no measurement of $\pi_+$ made,
the probability
of finding the particle at a location $x$ is given by
\begin{eqnarray} \label{4}
Prob(x\gets j ,t)=|A_{+}(x,j,t)+A_{-}(x,j,t)|^2\equiv \sum_{\mu,\mu'=+,-}\la j(t)|\hat \rho_{\mu\mu'}(x)|j(t)\ra,
\end{eqnarray}
where
\begin{eqnarray} \label{41}
\hat \rho_{\mu\mu'}(x)=\p_\mu \p_x\p_{\mu'}.
\end{eqnarray}
If a projective measurement of $\p_+$ is made, and a pointer records a result, $+$ or $-$, the interference is destroyed.
The probabilities of two possible outcomes, $(x,+)$ and $(x,-)$ are given by
\begin{eqnarray} \label{6}
Prob(x\gets \mu\gets j,t)=|A_{\mu}(x,j,t)|^2=\rho_{\mu\mu}(x,j,t),
\end{eqnarray}
where $\rho_{++}(x,t)$ and $\rho_{--}(x,t)$ are the probabilities of finding at $x$ a particle with positive and negative energy, respectively.
The total probability is conserved, $\int dx Prob(x,t) =\sum_{\mu=+,-}\int dx Prob(x,\mu,t)$, since the interference terms
$\rho_{+-}(x,t)$ and $\rho_{-+}(x,t)$ integrate to zero, $\int dx \rho_{+-}(x,t)= \int dx \rho_{-+}(x,t)=0$.
With several non-interacting particles present, one can use the probabilities in Eqs.(\ref{4}) to evaluate the energy resolved particle densities, as we will do next.
\section{$N$ distinguishable particles}
Consider next $N$ non-interacting distinguishable particles, each occupying one of the negative energy states (see Fig.2).

\begin{figure}
\includegraphics[angle=0,width=8cm]{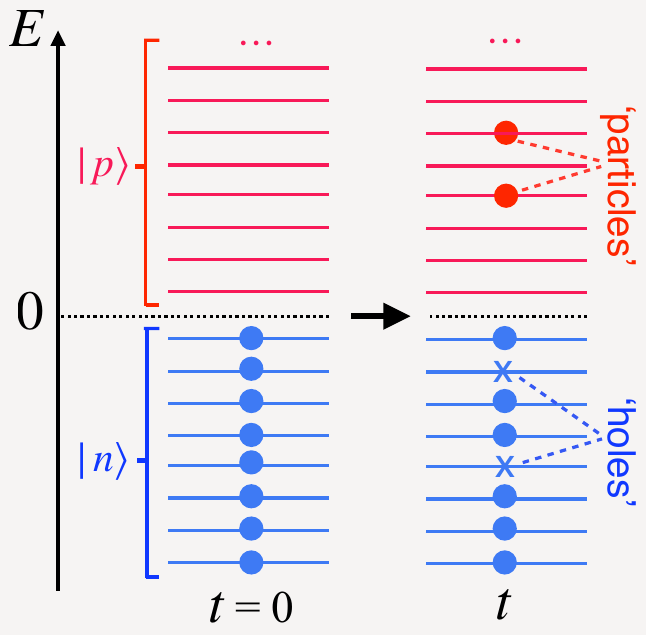}
\caption {a) Initial configuration of $N$ non-interacting distinguishable or identical particles occupying all negative
states $|n\ra$.
b) A time-dependent external field causes transitions, producing
\e{particles} with positive energies, and \e{holes} among the initially filled
negative energy levels.}
\label{FIG2}
\end{figure}

The Hamiltonian of the system is a sum,
\begin{eqnarray} \label{a-2}
\H^{dist}=\sum_{n=1}^n \h_{(n)}(t)
\end{eqnarray}
where $\h_{(n)}$ acts on the $n$-th particle's variable. With the first particle occupying the first of the negative states,
the second the second, and so on, the system's initial state, $|\Phi^{dist}(0)\ra=\prod_{n=1}^N|n\ra_{(n)}$, evolves into
\begin{eqnarray} \label{a-3}
|\Phi_N^{dist}(t)\ra=\prod_{n=1}^N|n(t)\ra_{(n)}.
\end{eqnarray}
Throughout the paper we will consider a one-particle density (occupation number), defined as the sum of probabilities
of finding each particle in a given state, regardless of where the remaining particles are. The density at $x$, $\rho(x,t)$, can, therefore,
be obtained as
\begin{eqnarray} \label{a-3b}
\rho(x,t)=\sum_{\mu,\mu'=+,-}\la \Phi_N^{dist}(t)| \hat \rho_{\mu\mu'}(x)|\Phi_N^{dist}(t)\ra\equiv \sum_{\mu,\mu'=+,-}\rho_{\mu\mu'}(x,t) ,
\end{eqnarray}
where
\begin{eqnarray} \label{a-3a}
\hat \rho_{\mu\mu'}(x)=\sum_{n=1}^N\p_\mu^{(n)} \p_x^{(n)}\p_{\mu'}^{(n)}.
\end{eqnarray}

As in Eqs.(\ref{a-2}) index $(n)$ indicates that an operator acts on the variables of the $n$-th particle.
In  the following we will assume that the sign of the particle's energy is made prior to measuring its position,
so that interference terms $\rho_{+-}(x,t)$ and  $\rho_{-+}(x,t)$ can be neglected,  $\rho(x,t)=\rho_{++}(x,t)+\rho_{--}(x,t)$,
and the densities of the positive- and negative-energy particles are given by
\begin{equation} \label{a1}
\begin{aligned}
\rho_{++}(x,t) &=\sum_n|\la x|\p_{+}|n(t)\ra|^2,\q \rho _{++}(x,0)=0,\\
\rho_{--}(x,t) &=\sum_n|\la x|\p_{-}|n(t)\ra|^2,\q \rho _{--}(x,0)=\sum_{n=1}^N |\la x|n\ra|^2.
\end{aligned}
\end{equation}
If  an extra particle in a superposition of, say, positive energy states,
$|\ph\ra = \sum_{p'} C_{p'} |p'\ra$, $\q \sum_{p'}|C_{p'}|^2=1,$
is added at $t=0$, the state of the system at $t$ is given by
$|\Phi_{N+1}^{dist}(t)\ra=|\ph(t)\ra_{(N+1)}\prod_{n=1}^N|n(t)\ra_{(n)}$.
where $|\ph(t)\ra=\u(t)|\ph\ra$.
The presence of an additional independent particle simply adds  to (\ref{a1}) an extra term
\begin{equation} \label{a3}
\begin{aligned}
\rho^\ph _{++}(x,t) &=\rho_{++}(x,t) + |\la x|\p_{+}|\ph(t)\ra|^2,\q  \rho^\ph _{++}(x,0)=|\la x|\ph\ra|^2,\\
\rho^\ph _{--}(x,t) &=\rho_{--}(x,t) + |\la x|\p_{-}|\ph(t)\ra|^2
, \q \rho^\ph _{--}(x,0)=\sum_n |\la x |n\ra|^2.
\end{aligned}
\end{equation}
 \subsection {\e{Particles} and \e{holes}}
For purely cosmetic purposes, one may refer to the positive energy particles as  \e{particles},
and treat the absence of a particle in a negative state as a \e{hole}.
Defined in this manner, the two densities become
\begin{equation} \label{a4}
\begin{aligned}
\rho^\ph_{ptcl}(x,t) &= \sum_n|\la x|\p_{+}|n(t)\ra|^2+|\la x|\p_{+}|\ph(t)\ra|^2 , \\
\rho^\ph_{hole}(x,t) &\equiv \rho^\ph_{--}(x,0)-\rho^\ph_{--}(x,t)= \sum_n |\la x |n\ra|^2-\sum_n|\la x|\p_{-}|n(t)\ra|^2- |\la x|\p_{-}|\ph(t)\ra|^2,
\end{aligned}
\end{equation}
so that $ \rho^\ph_{ptcl}(x,0)=|\la x|\ph\ra|^2$ and $\rho^\ph_{hole}(x,0)=0$.
Integration of Eqs.(\ref{a4}) yields gives the total number of \e{particles} and \e{holes} at time $t$,
\begin{equation} \label{a5}
\begin{aligned}
N^\ph_{ptcl}(t) &= \int \rho^\ph_{ptcle}(x,t)dx=\sum_{p,n}|\la p|n(t)\ra|^2 + |\la p| \ph(t)\ra|^2\\
N^\ph_{hole}(t) &=\int \rho^\ph_{hole}(x,t)dx= N-\sum_{n',n}|\la n'|n(t)\ra|^2 - |\la n|\ph(t)\ra|^2
\end{aligned}
\end{equation}
It is readily seen that at any time the number of \e{particles}  exceeds the number of \e{holes} by one, $N_{ptcl}(t)=N_{hole}(t)+1$.
{ Equations (\ref{a5}) have a simple interpretation. Scattering of the first $N$ particles into positive energy states creates both \e{particles}
and \e{holes}. Scattering of the extra particle into a negative state reduces the number of \e{holes}, hence the minus before the last term
in the second of Eqs.(\ref{a5}). Absolute square of the matrix element of the evolution operator,
$|\la n|\ph(t)\ra|^2=|\la n|\u(t)|\ph\ra|^2$ gives the corresponding probabilty.
\newline
 \subsection {Useful inequalities}
We note further that, with the chosen initial configuration (one particle per each negative state plus one more, a wave packet, built from positive states), the population of a positive energy level,
\begin{eqnarray} \label{a5a}
N^\ph_{ptcl}(p,t)\equiv \sum_{n=1}^{N+1} \la \Phi_{N+1}^{dist}(t)|\p^{(n)}_p|\Phi_{N+1}^{dist}(t)\ra =\sum_{n} |\la p|n(t)\ra|^2 -|\la p |\ph(t)\ra|^2, \q \p_p\equiv |p\ra\la p|,\q\q
\end{eqnarray}
cannot exceed unity.
Indeed,  one can introduce a new complete orthonormal set, $|\ph_p\ra$, spanning the positive sub-space and choose $|\ph_{N+1}\ra=|\ph\ra$.
Time-evolved states $|n(t)\ra=\u(t)|n\ra$ and $|\ph_p(t)\ra=\u(t)|\ph_p\ra$ now form a complete basis in the entire
Hilbert space, $\sum_n|n(t)\ra\la n(t)| + \sum_p|\ph_p(t)\ra\la \ph_p(t)|=\hat \1$. Writing
\begin{eqnarray} \label{a5b}
N^\ph_{ptcl}(p,t)=
\sum_n |\la n(t)|p\ra|^2 +|\la \ph(t)|p\ra|^2
\le \sum_n |\la n(t)|p\ra|^2 +\sum_p |\la \ph_p(t)|p\ra |^2 =\la p|p\ra=1\q\q
\end{eqnarray}
yields the desired result.
A similar argument shows that a \e{hole} occupation number
\begin{eqnarray} \label{a5c}
N^\ph_{hole}(n,t)\equiv 1- \sum_{n'=1}^{N+1} \la \Phi_{N+1}^{dist}(t)|\p^{(n')}_n|\Phi_{N+1}^{dist}(t)\ra
=1-\sum_{n'} |\la n'(t)|n\ra|^2 -|\la \ph(t)|p\ra|^2,\n
 \p_n\equiv |n\ra\la n|, \qq\q\q\q\q\q\q\q\q\q
\end{eqnarray}
in the second of Eqs.(\ref{a5})
always has a value between $0$ and $1$.

Next we see if anything would change in the case of identical particles obeying Fermi-Dirac or Bose-Einstein statistics.

\section{Non-relativistic fermions and bosons}
By Eq.(\ref{AA6}) of Appendix A, second-quantised versions of the Hamiltonian (\ref{1}) and the density operators (\ref{41}) are given by
\begin{equation} \label{2b}
\begin{aligned}
\H^{B,F}(t) &=\sum_{pp'}h_{pp'}\a^\+(p) \a(p')+\sum_{pn}h_{pn}\a^\+(p) \a(n)
+\sum_{np} h_{np}\a^\+(n)\a(p)+\sum_{nn'}\a^\+(n)\a(n'), \\
\hat \rho_{++}(x) &= \sum_{pp'}\a^\+(p)\a(p') \la p|x\ra\la x|p'\ra,\q \hat \rho_{+-}(x)= \sum_{pn}\a^\+(p)\a(n) \la p|x\ra\la x|n\ra \\
\hat \rho_{-+}(x) &= \sum_{np}\a^\+(n)\a(p) \la n|x\ra\la x|p\ra,\q \hat \rho_{--}(x)= \sum_{nn'}\a^\+(n)\a(n') \la n|x\ra\la x|n'\ra,   
\end{aligned}
\end{equation}
where the operators annihilating and creating a particle in a state $|j\ra$, $\a(j)$ and $\a^\+(j)$, obey
\begin{eqnarray} \label{1b}
[\a(j),\a^\+(j')\rx= \a(j)\a^\+(j')-\xi\a^\+(j')\a(j)=\delta_{jj'}, \q[\a^\+(j),\a^{\+}(j')\rx=[\a(j),\a(j')\rx=0, \q
\end{eqnarray}
with $\xi=1$ for bosons (B) , and $\xi=-1$ for fermions (F) (below we will write simply $[...]_\pm$ for $[...]_{\pm 1}$). Also the system's initial state becomes [cf. Eq.(\ref{AA2})]
\begin{eqnarray} \label{2ba}
|\Phi^{B,F}(0)\ra=\prod_{n=1}^N\a^\+(n)|\vc\ra,
\end{eqnarray}
where $|\vc\ra$ is the vacuum state, $\a(j)|\vc\ra=0$. The postive $\mu=+$ and negative $\mu=-$ energy densities are now given by
\begin{eqnarray} \label{1b1}
\rho_{\mu\mu}(x) = \la \Phi^{B,F}(t)|\hat\rho_{\mu\mu}(x)|\Phi^{B,F}(t)\ra =  \la \Phi^{B,F}(0)|\hat\rho_{\mu\mu}(x|t)|\Phi^{B,F}(0)\ra.
\end{eqnarray}
The first equality in Eq.(\ref{1b1}) corresponds to the Schr\"odinger picture, with time dependence incorporated in the wave function
$|\Phi^{B,F}(t)\ra=\exp\left [-i\int_0^t \H^{B,F}(t')dt'\right ] |\Phi^{B,F}(0)\ra$. In the second equality, time dependence is transferred to
operators in the Heisenberg representation,
\begin{eqnarray} \label{1b2}
\hat\rho_{\mu\mu'}(x|t)=\exp\left [i\int_0^t \H^{B,F}(t')dt'\right ]\hat\rho_{\mu\mu'}(x)\exp\left [-i\int_0^t \H^{B,F}(t')dt'\right ].
\end{eqnarray}
\subsection{Two kinds of time-dependent operators}
The time-evolved state can be obtained by first evolving the product state  [cf. (\ref{a-3})], and then symmetrising the result.
This yields
\begin{eqnarray} \label{2b1}
|\Phi^{B,F}(t)\ra=\prod_{n=1}^N\a^+(n(t))|\vc\ra,
\end{eqnarray}
where the newly introduced operators
$\a^\+(j(t))$, $j=p,n$, create a particle in a time-evolved state
$|j(t)\ra=\u(t)|j\ra$. They obey the same commutation rules \cite{FeynS}, since the evolution of one-particle states is unitary,
and $|j(t)\ra$ also form a complete orthonormal basis,
\begin{eqnarray} \label{3b}
[\a(j(t)),\a^\+(j'(t))\rx=\la j(t)|j'(t)\ra=\la j|j'\ra=\delta_{jj'}.
\end{eqnarray}
Since $|j(t)\ra=\sum_{j'} |j'\ra \la j'|\u(t)|j\ra$, one finds [ cf. (\ref{AA5}) of Appendix A]
\begin{eqnarray} \label{4b}
\a(j(t))=\sum_{j'}  \la j'|\u(t)|j\ra^*\a(j')=\sum_{j'}  \la j|\u(-t)|j'\ra\a(j').
\end{eqnarray}
This can be inverted
to yield an expansion
\begin{eqnarray} \label{4b}
\a(j)=\sum_{j'} { \la j'|}\u(-t)|j\ra^*\a(j'(t))=\sum_{j'}  \la j|\u(t)|j'\ra\a(j'(t)),
\end{eqnarray}
more useful for our purposes.
We note that that $\a^\+(j(t))$ should not be confused with the Heisenberg operators $\a(j|t)$,
 which satisfy
\begin{eqnarray} \label{4b1}
\partial_t \a(j|t)=i[\H^{B,F}(t),\a(j|t)]_+,
\end{eqnarray}
 and obey the same commutation relations (\ref{1b}).
 Time dependent operators of both types will be useful in what follows.
\subsection{\e {Particle-hole} production}
Using the first equality in Eq.(\ref{1b1}), the time-evolved state (\ref{2b1}), and expressing  $\a^\+(j)$ in terms of $\a^\+(j(t))$ with the help of  Eq.(\ref{4b}), one easily finds
\begin{equation} \label{8b1}
\begin{aligned}
\rho_{ptcl}(x,t)&=\la \Phi^{B,F}(t)|\hat \rho^{B,F}_{++}(x)|\Phi^{B,F}(t)\ra = \sum_{n}|\la x|\hat \pi_+|n(t)\ra|^2 \\
\rho_{hole}(x,t)&=\la \Phi^{B,F}(0)|\hat \rho^{B,F}_{--}(x)|\Phi^{B,F}(0)\ra-\la \Phi^{B,F}(t)|\hat \rho^{B,F}_{--}(x)|\Phi^{B,F}(t)\ra \\
&=\sum_{n}\left [|\la x|n\ra|^2-|\la x|\p_{-}|n(t)\ra|^2\right].
\end{aligned}
\end{equation}
Equations (\ref{8b1}) have the same form for bosons and fermions, and coincide with Eqs. (\ref{a3}) obtained earlier for distinguishable particles.
The search for statistical effects in one-particle densities must continue elsewhere.
\subsection{\e {Particle-hole} production with an extra \e{particle} present}
One may expect that adding at $t=0$ an extra particle  in a state $|\ph\ra = \sum_{p'} C_{p'} |p'\ra$ might affect the
\e {particle-hole} production for bosons and fermions differently.
Indeed, a hand waving argument suggests that with some of the \e{particle} states
already occupied, fermions would be less likely to leave negative energy states
due to Pauli principle. For bosons, with their tendency to \e{bunch together} \cite{bunch}, the
 transitions could be enhanced by the presence of the additional \e{particle}.
It is, however, easy to  demonstrate that this is not the case.
\newline
The state of the $(N+1)-$particle system at $t$  is now given by
\begin{eqnarray} \label{11b}
|\Phi_\ph^{B,F}(t)\ra=\a^\+(\ph(t))\prod_{n=1}^N\a^\+(n(t))|\vc\ra=\a^\+(\ph(t))|\Phi^{B,F}(t)\ra ,
\end{eqnarray}
where $\a^\+(\ph(t))$ creates a particle in a state $|\ph(t)\ra=\u(t)|\ph\ra=\sum_{p}C_p|p(t)\ra$.
To analyse the effect of adding an extra \e{particle} it is
convenient to rewrite operators $\hat \rho_{\mu\mu}(x)$ in Eqs.(\ref{2b}) in an equivalent form,
\begin{eqnarray} \label{12b1}
\hat \rho_{\mu\mu}(x)=\a_\mu^\+(x)\a_\mu(x), \q \a_{+} (x)= \sum_{p} \la x|p\ra\a(p), \q \a_{-} (x)= \sum_{n} \la x|n\ra\a(n).
\end{eqnarray}
 For the \e{particle} density we, therefore, have
\begin{eqnarray} \label{13b}
\rho^\ph_{ptcl}(x,t)= \la \Phi^{B,F}(t)|\a(\ph(t))\a^\+_+(x)\a_+(x) \a^\+(\ph(t))|\Phi^{B,F}(t)\ra,\q\q\q\q
\end{eqnarray}
where, [cf. Eq.(\ref{12b1})]
\begin{equation} \label{14b}
  \begin{aligned}
\a(\ph(t))\a^\+_+(x)&=\xi \a^\+_+(x)\a(\ph(t)) + \la x|\p_+|\ph(t)\ra^* \\
\a_+(x)\a^\+(\ph(t))&= \left (\a(\ph(t))\a^\+_+(x)\right )^\+=\xi \a^\+(\ph(t))\a_+(x) +  \la x|\p_+|\ph(t)\ra.
\end{aligned}
\end{equation}
Inserting (\ref{14b}) into (\ref{13b}), one  notes that the terms containing $\xi$ vanish since $\a(\phi(t))|\Phi^{B,F}(t)\ra=0$.
The term proportional to $\xi^2=1$ yields the density  without an extra \e{particle} in Eq.(\ref{8b1}).
Indeed, interchanging $\a(\ph(t))$ and $\a^\+(\ph(t))$ gives [cf. Eq.(\ref{AA3})]
\begin{eqnarray} \label{14b}
\la \Phi^{B,F}(t)|\a^\+_+(x)[1 +\xi \a^\+(\ph(t))\a(\ph(t)]\a_+(x) )|\Phi^{B,F}(t)\ra= \rho_{ptcl}(x,t),
\end{eqnarray}
where we have used $\a(\ph(t))\a_+(x)|\Phi^{B,F}(t)\ra=\xi \a_+(x) \a(\ph(t) )|\Phi^{B,F}(t)\ra=0$.
After repeating the calculation also for \e{holes} one finds
\begin{equation} \label{10b}
  \begin{aligned}
\rho^\ph_{ptcl}(x,t) &=\sum_{n}|\la x|\hat \pi_+|n(t)\ra|^2+|\la x|\p_{+}|\ph(t)\ra|^2,\\
\rho^\ph_{hole}(x,t) &=\sum_n |\la x |n\ra|^2- \sum_n|\la x|\p_{-}|n(t)\ra|^2- |\la x|\p_{-}|\ph(t)\ra|^2,
\end{aligned}
\end{equation}
which is the result (\ref{a4}), obtained earlier for distinguishable particles.
We note that, for both fermions and bosons,
 Eqs.(\ref{10b})
can be rewritten in a more symmetric form (see Appendix C),
\begin{equation} \label{10b1}
  \begin{aligned}
\rho^\ph_{ptcl}(x,t) &=\sum_{n}|\la x|\hat \pi_+|n(t)\ra|^2+|\la x|\p_{+}|\ph(t)\ra|^2,\\
\rho^\ph_{hole}(x,t) &=\sum_p|\la x|\p_{-}|p(t)\ra|^2- |\la x|\p_{-}|\ph(t)\ra|^2.
\end{aligned}
\end{equation}
Finally, evaluation of \e{particle} and \e{hole} occupation numbers,
\begin{equation} \label{10b2}
  \begin{aligned}
N^\ph_{ptcl}(p,t)&= \la\Phi_\ph^{B,F}(t)|\a^\+(p)\a(p)|\Phi_\ph^{B,F}(t)\ra,\\
N^\ph_{hole}(n,t)&=1- \la\Phi_\ph^{B,F}(t)|\a^\+(n)\a(n)|\Phi_\ph^{B,F}(t)\ra,
\end{aligned}
\end{equation}
also yields Eqs.(\ref{a5a}) and (\ref{a5c}) obtained earlier in Sect. III.
The fact that both for bosons and fermions $N^\ph_{ptcl}(p,t)$  may not exceed unity [cf. Sect.IIIB] is, therefore, a property of the initial configuration (one particle per each state of an orthogonal basis), rather than a consequence of the Pauli exclusion principle. Next we consider relativistic fermions.
\section{Dirac electrons in one dimension}
Analysis of the previous Section can be extended to relativistic spin-$1/2$ fermions with only minor changes.
One-particle Dirac (D) equation in one spatial dimension takes the form \cite{FV} ($c=1, \hbar=1$),
\begin{eqnarray} \label{1c}
i\partial_t |\psi\ra =\h^D|\psi\ra=[\h^D_0+\hat V(x,t)]|\psi\ra.
\end{eqnarray}
 In the force-free case $\hat V=0$, the spectrum consists  of positive (+) and negative (-)  continua.
For simplicity we consider a system confined to a finite volume, so the spectrum is discrete,
\begin{eqnarray} \label{1c2}
\h^D_0|\phi^\pm_k\ra= \pm E_k |\phi^\pm_k\ra, \q E_k=\sqrt{\alpha^2(k-1)^2+m^2}>0,
\end{eqnarray}
where $m$ is  the rest mass of an electron, $\alpha$ is a constant which depends on the confinement, $k$ numbers the states, as shown in Fig.\ref{FIG3}.
We have also abandoned earlier notations $|n\ra$ and $|p\ra$ in favour of more convenient
$|\pp\ra$ and $|\nm\ra$,
\begin{eqnarray} \label{2c}
|\phi^\pm_k\ra = |v^\pm_k\ra|k\ra, \q \la k|k'\ra=\delta_{kk'}, \q \la v^{\mu}_k|v_k^{\mu'}\ra=\delta_{\mu\mu'}, \q \mu=+,-, \q
\end{eqnarray}
 which form a complete orthonormal basis in a Hilbert space $\mathcal H =\mathcal H_2\otimes \mathcal H_\infty$, a direct product of a two-dimensional
$\mathcal H_2$ spanned by any pair $|v^\pm_k\ra$, and an infinite-dimensional $\mathcal H_\infty$, spanned by the basis of momentum states $|k\ra$,
 or of position states $|x\ra=\sum_k |k\ra\la k|x\ra$.
The completeness relation reads
\begin{eqnarray} \label{3c}
\sum_\mu \sum_k | \phi_k^\mu \ra \la \phi_k^\mu|=\sum_k |\pp\ra \la \pp|+\sum_k |\nm\ra \la \nm|=\p_++\p_- =\hat \1,
\end{eqnarray}
and the one-particle Hamiltonian is defined by its matrix elements,
\begin{eqnarray} \label{2c1}
\h^{D}= \sum_{\mu,\mu'}\sum_{k,k'}|\phi^\mu _{k}\ra  h_{kk'}^{\mu\mu'} \la \phi^{\mu'}_{k'}|=\sum_k E_k [|\phi^+_{k}\ra\la \phi^+_k| -  |\phi^-_{k}\ra\la \phi^-_k|]+ \sum_{\mu,\mu'}\sum_{k,k'}|\phi^\mu _{k}\ra  V_{kk'}^{\mu\mu'} \la \phi^{\mu'}_{k'}|,\q\q
\end{eqnarray}
where $h_{kk'}^{\mu\mu'}(x)\equiv \la \phi^\mu _{k}|\h^D| \phi^{\mu'}_{k'}\ra$ and $V_{kk'}^{\mu\mu'}(x)\equiv \la \phi^\mu _{k}|\hat V| \phi^{\mu'}_{k'}\ra=\la v_k^\mu|v_{k'}^{\mu'}\ra\la k|\hat V|k'\ra=V_{k'k}^{\mu'\mu*}$.
Multiplying
$\p_x = |x\ra \la x|$ by unity (\ref{3c})
yields [cf. Eq.(\ref{a-3a})]
\begin{eqnarray} \label{5c}
\hat \rho(x)= \sum_{\mu\mu'}\hat \rho_{\mu\mu'}(x) =\sum_{\mu,\mu'}\p_\mu \p_x\p_{\mu'}=\sum_{\mu,\mu'}\sum_{kk'}|\phi^\mu_k\ra \la v_k^\mu|v_{k'}^{\mu'}\ra\la k|x\ra\la x|k'\ra\la \phi^{\mu'}_{k'}|
\end{eqnarray}
\subsection{Second quantisation}
As in the non-relativistic case, the second-quantised versions of the Hamiltonian  (\ref{2c1}) and the density operators $\hat \rho_{\mu\mu'}(x)$ in Eq.(\ref{3c})
are readily obtained by replacing bras and kets by the corresponding annihilation and creation operators
which now obey fermionic commutation relations. We, therefore, have
 \begin{equation} \label{6c}
   \begin{aligned}
 \H^D(t) &=\sum_{\mu\mu'}\sum_{k,k'} h^{\mu\mu'}_{kk'} \a^\+( \phi^\mu_k)\a( \phi^{\mu'}_{k'}),\\
\hat  \rho_{\mu\mu'}(x)&=\sum_{kk'} \la v_k^\mu|v_{k'}^{\mu'}\ra\la k|x\ra\la x|k'\ra\a^\+( \phi^\mu_k)\a( \phi^{\mu'}_{k'}),\\
 [\a(\phi^\mu_k),&\a^\+({ \phi^{\mu'}_{k'}})]_- =\la \phi_k^\mu| \phi_{k'}^{\mu'}\ra = \delta_{\mu\mu'}  \delta_{kk'}.
\end{aligned}
\end{equation}
The vacuum (ground) state of the system can be obtained by filling all the  negative energy states (the Fermi sea) \cite{Greiner}
 \begin{eqnarray} \label{7c}
|\text{Vac}\ra=\prod_k \a^\+(\phi^-_k)|\vac\ra, \q \a^\+(\phi^\mu_k)|\vac\ra=0.
\end{eqnarray}

\subsection{ {Electron-positron} production with an extra {electron} present}
With all prerequisites in place, one can simply repeat the calculation of the previous Section,
assuming that at $t=0$ there is a single positive-energy electron in a state $|\ph\ra= \sum_{k'}C_{k'}|\phi^+_{k'}\ra$.
\begin{figure}
\includegraphics[angle=0,width=8cm]{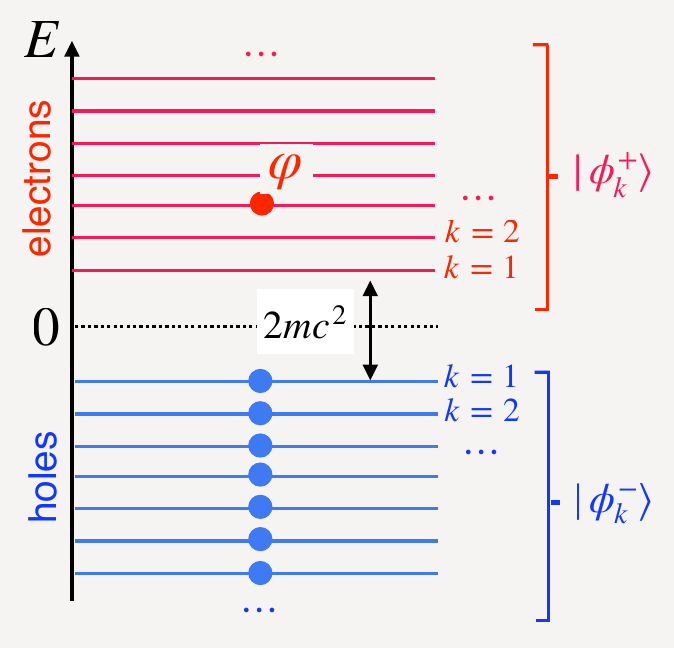}
\caption {Negative and positive energy continua for relativistic fermions and bosons.
For the fermions, the ground state can be constructed by filling the negative energy states
of the {Fermi sea}. For bosons no such construction is possible.
Extra particle is added in a positive energy state (more generally, in a superposition of such states).
Once the field causing transitions between the state is switched off,
we wish to evaluate spatially resolved densities of electrons and holes (positrons),
in the case of fermions, and of bosons and anti-bosons in  the case of bosons.}
\label{FIG3}
\end{figure}

The initial state
\begin{eqnarray} \label{7c}
|\Phi^D_\ph (0)\ra =\a^\+(\ph)|\text{Vac}\ra
\end{eqnarray}
evolves into
$|\Phi^D_\ph (t)\ra =\a^\+(\ph(t))\prod_k \a^\+(\phi^-_{k}(t)|\vac\ra$.
Unitary evolution conserves scalar products, and the time-evolved one-particle states $|\phi^\mu_{k}(t)\ra$
form a complete orthonormal basis,
$\sum_\mu \sum_k | \phi_k^\mu(t) \ra \la \phi_k^\mu(t)|=\hat \1$.
Repeating the steps of Sect.IVC, after renaming \e{particles} and \e{holes} as (el)ectrons and (pos)itrons  \cite{Greiner},
one recovers  Eqs.(\ref{10b}) for the corresponding densities,
\begin{equation} \label{10c}
  \begin{aligned}
\rho^\ph_{el}(x,t) &=\sum_{k}\la \nm(t)|\hat \pi_+| x \ra \la x|\hat \pi_+|\nm(t)\ra+\la \ph(t)|\hat \pi_+| x \ra \la x|\hat \pi_+|\ph(t)\ra,\\
\rho^\ph_{pos}(x,t) &=\sum_{k}\left [\la \nm| x \ra \la x|\nm\ra- \la \nm(t)|\hat \pi_-| x \ra \la x|\hat \pi_-|\nm(t)\ra\right ]- \la \ph(t)|\hat \pi_-| x \ra \la x|\hat \pi_-|\ph(t)\ra.
\end{aligned}
\end{equation}
Note that  now $\la \nm(t)|\hat \pi_\mu| x \ra \la x|\hat \pi_\mu|\nm(t)\ra$ is no longer given by  $|\la x|\hat \pi_\mu|\nm(t)\ra|^2$.
However, one can still use the completeness relation (\ref{3c}) to cast Eqs.(\ref{10c}) into  a symmetric form, similar to Eqs.(\ref{10b1}) (see Appendix C),
\begin{equation} \label{10c1}
  \begin{aligned}
\rho^\ph_{el}(x,t) &=\sum_{k}\la \nm(t)|\hat \pi_+| x \ra \la x|\hat \pi_+|\nm(t)\ra+\la \ph(t)|\hat \pi_+| x \ra \la x|\hat \pi_+|\ph(t)\ra,\\
\rho^\ph_{pos}(x,t) &=\sum_{k} \la \pp(t)|\hat \pi_-| x \ra \la x|\hat \pi_-|\pp(t)\ra- \la \ph(t)|\hat \pi_-| x \ra \la x|\hat \pi_-|\ph(t)\ra,\\
\end{aligned}
\end{equation}
and avoid any explicit mention of the filled Fermi sea.
\newline
The advantage of calculating the densities in this manner is twofold.
Firstly, there is no need to solve Heisenberg equations of motion, since the matrix elements of one-particle evolution operator
are readily obtained by transforming to the evolved basis $|\ph^\mu_k\ra  \to |\ph^\mu_k(t)\ra$.
Secondly, and more importantly, derivation of (\ref{10c})  does not rely on the Pauli principle.
The same densities would be for an initial state placing one boson per each negative state and adding one boson in a wave packet
built from only positive energy states. However, the method is not general, and as will be shown shortly, cannot be extended to relativistic bosons.
\subsection{Alternative derivation}
Alternatively, one can rely on the second equality in Eq.(\ref{1b}) and evaluate the operators $\a(\phi_k^\mu)$ in their Heisenberg form,
 $\a(\phi_k^\mu|t)$.
 It is customary to rename
$\a(\nm) \to \b^\+(\nm)$ and $\a^\+(\nm) \to \b(\nm)$, $\b(\nm)|\text{Vac}\ra =0$,
upon which operator $\hat  \rho_{--}(x)$ in Eq.(\ref{6c}) takes the form
\begin{eqnarray} \label{0d}
\hat  \rho_{--}(x)=\sum_{k}\la k|x\ra\la x|k\ra  -\sum_{kk'} \la v_k^-|v_{k'}^{-}\ra\la k|x\ra\la x|k'\ra b^\+( \phi^{-}_{k'})\b^( \phi^-_k).
\end{eqnarray}
 Integrating (\ref{0d})  over $dx$ shows that
 \begin{eqnarray} \label{1d}
\hat\rho_{pos}(x) \equiv \sum_{kk'} \la v_k^-|v_{k'}^{-}\ra\la k|x\ra\la x|k'\ra b^\+( \phi^{-}_{k'})\b( \phi^-_k)
\end{eqnarray}
 describes the positron (hole) density.
In the new notations, the Hamiltonian in (\ref{6c}) becomes
\begin{equation} \label{2d}
\begin{aligned}
\H^D(t) =
& \sum_{k,k'} h^{++}_{kk'} \a^\+( \phi^+_k)\a( \phi^{+}_{k'}) +\sum_{k,k'} h^{+-}_{kk'}(t) \a^\+( \phi^+_k)\b^\+( \phi^{-}_{k'})+\\
&\sum_{k,k'} h^{-+}_{kk'} \a( \phi^+_k)\b ( \phi^{-}_{k'})-\sum_{k,k'} h^{--}_{k'k}(t) \b^\+( \phi^-_k)\b(\phi^{-}_{k'}),
\end{aligned}
\end{equation}
where we omitted an infinite  $c$-number term $\sum_k h^{--}_{kk}(t)$.
With $\a^\+(\phi^+_k)$ and $\b^\+(\phi^-_k)$ interpreted as operators creating electrons and positrons, respectively,
$h^{+-}_{kk'}$ and $h^{-+}_{kk'}$ become the amplitudes of creation and destruction  of electron-positron pairs \cite{FeynS}.
Note that the force-free ($\hat V=0$) Hamiltonian $\H_0^D(t)= \sum_k E_k [\a^\+( \phi^+_k)\a( \phi^{+}_{k})+ \b^\+( \phi^-_k)\b( \phi^{-}_{k})]$, $E_k>0$,
describes free electrons and positrons both having positive energies.
\newline
Evaluating the commutators $[\a( \phi^+_k),\H^D]_+$ and $[\b^\+( \phi^-_k),\H^D]_+$ with the help of an identity
 $[\hat A,\hat B \hat C]_+=[\hat A,\hat B]_-\hat C-\hat B [\hat A,\hat C]_-$
yields the
equations of motion
 \begin{eqnarray} \label{3d}
  \begin{cases}
 i\partial_t \a( \phi^+_{k}|t) = [ \a( \phi^+_{k}|t),\H^{D} ]_+= \sum_{k'} h^{++}_{kk'} \a( \phi^+_{k'}|t)+ \sum_{k'}  h^{+-}_{kk'}\b^\+( \phi^-_{k'}|t), \\
 i\partial_t \b^\+( \phi^-_{k}|t) = [{\b^\+( \phi^-_{k}|t)},\H^{D} ]_+ = \sum_{k'}  h^{--}_{kk'}\b^\+( \phi^-_{k'}|t)+ \sum_{k'} h^{-+}_{kk'} \a( \phi^+_{k'}|t).
   \end{cases}
\end{eqnarray}
The matrix in the r.h.s. of Eq.(\ref{3d}) is hermitian, which guarantees conservation of particles,
expressed in the old notations as   $\sum_{\mu}\sum_k  \a^\+( \phi^\mu_{k}|t) \a( \phi^\mu_{k}|t)=const$ .
\newline
It is easy to check that Eqs.(\ref{3d}) have a simple solution
relating Heisenberg operators to their Schr\"{o}dinger counterparts as
 \begin{equation} \label{5d}
  \begin{aligned}
 \a( \phi^+_{k}|t) &= \sum_{k'} \la \phi^+_k|\phi^+_{k'}(t)\ra  \a( \phi^+_{k'})+ \sum_{k'}\la \phi^+_k|\phi^-_{k'}(t)\ra\b^\+( \phi^-_{k'}),\q  \a( \phi^+_{k}|0)=\a( \phi^+_{k}), \\
 \b^\+( \phi^-_{k}|t) &= \sum_{k'} \la \phi^-_k|\phi^+_{k'}(t)\ra \a( \phi^+_{k'})+ \sum_{k'}  \la \phi^-_k|\phi^-_{k'}(t)\ra\b^\+( \phi^-_{k'}) ,\q  \b^\+( \phi^-_{k}|0)=\b^\+( \phi^-_{k}),
\end{aligned}
\end{equation}
where $|\phi^\mu_{k'}(t)\ra$ satisfies the one-particle Dirac equation (\ref{1c}) with the initial condition $|\phi^\mu_{k'}(0)\ra=|\phi^\mu_{k'}\ra$ (see Appendix D).
Inserting  Eq.(\ref{5d}) into the expressions
 \begin{equation} \label{5d1}
   \begin{aligned}
\hat \rho_{el}(x|t) &= \sum_{kk'} \la v_k^+|v_{k'}^{+}\ra\la k|x\ra\la x|k'\ra a^\+( \phi^{+}_{k}|t)\a( \phi^+_{k'}|t), \\
\hat\rho_{pos}(x|t) &=\sum_{kk'} \la v_k^-|v_{k'}^{-}\ra\la k|x\ra\la x|k'\ra b^\+( \phi^{-}_{k'}|t)\b( \phi^-_k|t),
\end{aligned}
\end{equation}
and evaluating $\la \Phi_\ph^{D}(0)|\hat \rho^D_{el}(x|t)| \Phi_\ph^{D}(0)\ra$  and $\la \Phi_\ph^{D}(0)|\hat \rho^D_{pos}(x|t)| \Phi_\ph^{D}(0)\ra$,
one recovers equations (\ref{10c})-(\ref{10c1}) for spatially resolved electron and positron densities.
\newline
Although somewhat more cumbersome than the method of Sect. IVA, the approach based on Heisenberg equations of motion is more general.
Next we apply it to study the case of relativistic bosons.
\section{Klein-Gordon bosons in one dimension}
A similar treatment can be extended to spinless bosons, albeit only up to a point.
In its Hamiltonian form \cite{Greiner}, one-particle Klein-Gordon (KG) equation in one spatial dimension is given by
\begin{eqnarray} \label{1e}
i\partial_ t|\psi\ra =\h^{KG}|\psi\ra=[\h^{KG}_0+\hat V(x,t)]|\psi\ra.
\end{eqnarray}
As in the case of Dirac electrons, the spectrum of the field-free Hamiltonian consists of the positive and negative parts [cf. Eq.(\ref{1c})]
\begin{eqnarray} \label{2e}
\h^\K_0|\phi^\pm_k\ra= \pm E_k |\phi^\pm_k\ra, \q E_k=\sqrt{\alpha^2(k-1)^2+m^2}>0,
\end{eqnarray}
with the eigenstates
\begin{eqnarray} \label{3e}
|\phi^\pm_{k}\ra = |v_k^\pm\ra |k\ra, \q \la k|k'\ra=\delta_{kk'}, \q \la v^{+}_k|v_k^{+}\ra=- \la v^{-}_k|v_k^{-}\ra=1, \q  \la v^{+}_k|v_k^{-}\ra= \la v^{-}_k|v_k^{+}\ra=0,\q
\end{eqnarray}
subjected to an unusual \e{scalar product},
\begin{eqnarray} \label{4e}
\la \phi^+_k|\phi^+_{k'}\ra = -\la \phi^-_k|\phi^-_{k'}\ra=\delta_{kk'},\q \la \phi^\pm_k|\phi^\mp_{k'}\ra=0.
\end{eqnarray}
An arbitrary state $|\psi\ra$ can be expanded as
$|\psi\ra=\sum_\mu \sum_k A_{\mu k} |\phi^\mu_{k}\ra$,
and the completeness relation is seen to be given by
\begin{eqnarray} \label{5e}
\sum_k |\phi^+_{k}\ra\la \phi^+_k|-\sum_k |\phi^-_{k}\ra\la \phi^-_k|\equiv \p_+-\p_-=\hat \1.
\end{eqnarray}
Multiplying one-particle operator $\p_x=|x\ra\la x|$ on both sides by the unity (\ref{5e})
yields
\begin{eqnarray} \label{6e}
\rhoo^\K (x)=\sum_{\mu\mu'}\lm_{\mu\mu'}\rhoo_{\mu\mu'}(x)=
\sum_{\mu,\mu'}\lm_{\mu\mu'}\sum_{kk'}|\phi^\mu_k\ra \la v_k^\mu|v_{k'}^{\mu'}\ra\la k|x\ra\la x|k'\ra\la \phi^{\mu'}_{k'}|\n
\end{eqnarray}
where $\lm_{\mu\mu'} =1$ if $\mu=\mu'$, $-1$ if $\mu\ne \mu'$.
\newline
In a similar manner, for the Hamiltonian (\ref{1e}) we find
\begin{equation} \label{7e}
  \begin{aligned}
\h^{KG} & =
 \sum_{\mu,\mu'}\lm_{\mu\mu'}\sum_{k,k'}|\phi^\mu _{k}\ra  h_{kk'}^{\mu\mu'} \la \phi^{\mu'}_{k'}|,\\
h_{kk'}^{\mu\mu'}(x) & \equiv \la \phi^\mu _{k}|\h^\K| \phi^{\mu'}_{k'}\ra= |E_k|\delta_{\mu\mu'}\delta_{kk'}+\la v^\mu_k|v^{\mu'}_{k'}\ra \la k|\hat V(x)|k'\ra,
\end{aligned}
\end{equation}
where we have used
$\la \phi^-_k|\h_0^{KG}|\phi^-_k\ra = -E_k\la \phi^-_k|\phi^-_k\ra= |E_k|$.
\subsection{Second quantisation}
As in the case of Dirac electrons, the second-quantised versions of the Hamiltonian  (\ref{7e}) and the density operators $\rhoo_{\mu\mu'}(x)$ in Eq.(\ref{6e})
are readily obtained by replacing bras and kets by the corresponding annihilation and creation operators
which now obey somewhat unusual commutation relations [cf. Eqs.(\ref{AA3}) of Appendix A]. We, therefore, have
 \begin{equation} \label{8e}
   \begin{aligned}
 \H^\K(t) &=\sum_{\mu\mu'}\lm_{\mu\mu'}\sum_{k,k'} h^{\mu\mu'}_{kk'} \a^\+( \phi^\mu_k)\a( \phi^{\mu'}_{k'}),\\
\hat  \rho_{\mu\mu'}(x) & =\sum_{kk'} \rho^{\mu\mu'}_{kk'}(x) \a^\+( \phi^\mu_k)\a( \phi^{\mu'}_{k'})=\sum_{kk'} \la v_k^\mu|v_{k'}^{\mu'}\ra\la k|x\ra\la x|k'\ra\a^\+( \phi^\mu_k)\a( \phi^{\mu'}_{k'}),\\
[\a(\phi^+_k), &\a^\+(\phi^+_{k'})]_+=\la\phi^+_k  | \phi^+_{k'}\ra =\delta_{kk'}, \q [\a(\phi^-_k), \a^\+(\phi^-_{k'})]_+=\la\phi^-_k  | \phi^-_{k'}\ra=-\delta_{kk'}.
\end{aligned}
\end{equation}
Renaming $\a(\phi^-_k) \to \b^\+(\phi^-_k)$, $\a^\+(\phi^-_k) \to \b(\phi^-_k)$ restores the usual
bosonic commutation rules,
\begin{eqnarray} \label{9e}
[\a(\phi^+_k), \a^\+(\phi^+_{k'})]_+ =\delta_{kk'}, \q [\b(\phi^-_k), \b^\+(\phi^-_{k'})]_+=\delta_{kk'}.
\end{eqnarray}
Expressed in terms of the new operators, the Hamiltonian has a form similar to that of $\H^{D}$ in Eq.(\ref{2d}),
\begin{equation} \label{10e}
\H^{KG}(t) =
\sum_{k,k'} \left [ h_{kk'}^{++}\a^\+(\phi^+ _{k})  \a (\phi^{+}_{k'})-h_{kk'}^{+-} \a^\+(\phi^+ _{k})   \b^\+ (\phi^{-}_{k'}) \\
  -  h_{kk'}^{-+} \b(\phi^- _{k})  \a (\phi^{+}_{k'}) +h_{kk'}^{--}\b^\+(\phi^+ _{k'})   \b (\phi^{+}_{k}) \right ] ,
\end{equation}
where we have omitted a $c$-number term $\sum_k h_{kk}^{--}$ arising from interchanging $\b(\phi^-_{k})$ and $\b^\+(\phi^-_{k})$
in last term of Eq.(\ref{10e}).
The Hamiltonian (\ref{10e}) can be seen as describing two types of bosons, corresponding to commuting operators $\a (\phi^{+}_{k})$ and $\b(\phi^{+}_{k})$,
which appear and disappear in pairs. We will refer to them as bosons  (bos) and anti-bosons (abos), respectively.
The number operators for both types of particles are given by the usual expressions,
$\hat N_{bos}= \sum_k \a^\+(\phi^+_k)\a(\phi^+_k)$ and $\hat N_{abos}= \sum_k \b^\+(\phi^-_k)\b(\phi^-_k)$,
and it is a simple matter to check that the difference $\hat N_{bos}-\hat N_{abos}$ is maintained constant at all times.
\newline
This is as far as one can follow the construction of the Fock space for Dirac fermions in the case of relativistic spinless bosons.
Indeed, to complete the description of bosons created and annihilated in pairs by the external field $V$, one requires a vacuum state, such that
\begin{eqnarray} \label{12e}
\a(\phi^+_{k})|\text{Vac}\ra_\K = \b(\phi^-_{k})|\text{Vac}\ra_\K=0.
\end{eqnarray}
For electrons such a state could be constructed simply by filling up all negative levels of the Fermi sea. For bosons, one cannot construct $|\text{Vac}\ra$
by acting with operators $\a^\+(\phi^\mu_{k})$ on the original vacuum state of the second-quantised theory, $|\text{vac}\ra$. Equation (\ref{12e}) needs to be postulated, in a departure from the
usual second quantisation procedure \cite{FeynS}. There is no simple expression  for the time-evolved vacuum state, and one has no option but to follow the steps of Sect.VC.
\newline

 \subsection{ {Boson-antiboson production} production with an extra {boson} present}
 As before, we consider the case where at $t=0$ there is a single boson present in a superposition (wave packet) of positive energy states,
 $|\ph\ra=\sum_{k} C_{k}|\phi^+_{k}\ra$, and the initial state is given by
 \begin{eqnarray} \label{13e}
|\Phi^\K_\ph (0)\ra =\a^\+(\ph)|\text{Vac}\ra_\K.
\end{eqnarray}
It remains to explore the physical significance of the second-quantised version of the operators $\rhoo_{\mu\mu'}(x)$ defined in Eq.(\ref{6c}), in the new notations given by
\begin{equation} \label{11e}
  \begin{aligned}
\rhoo_{++}(x) &=  \sum_{k,k'}  \rho_{kk'}^{++}(x)\a^\+(\phi^+_{k})\a(\phi^+_{k'}),\\
\rhoo_{+-}(x) &=  \sum_{k,k'}  \rho_{kk'}^{+-}(x)\a^\+({\phi^+_{k}}){ \b^\+}(\phi^-_{k'})=\rhoo^\+_{-+}(x) ,\\
\rhoo_{--}(x) &=  \sum_{k,k'}  \rho_{k'k}^{--}(x)\b^\+(\phi^-_{k})\b(\phi^-_{k'})+\sum_k  \rho_{kk}^{--}(x).
\end{aligned}
\end{equation}
Integration of Eqs.(\ref{11e}) yields [cf. Eqs.(\ref{3e})]
\begin{eqnarray} \label{15e}
\int\rhoo_{++}(x)dx= \hat N_{bos}, \q \int\rhoo_{--}(x)dx= -\hat N_{abos}-\sum_k 1,
\end{eqnarray}
while $\int\rhoo_{+-}(x)dx=  \int\rhoo_{-+}(x)dx= 0$.
By treating bosons and anti-bosons as two different kinds of particles \cite{Schw}, one is able to omit the terms $\rhoo_{+-}(x)$ and $\rhoo_{+-}(x)$.
The operator
$\hat q_{abos}(x) \equiv \sum_{k,k'}  \rho_{k'k}^{--}(x)\b^\+(\phi^-_{k})\b(\phi^-_{k'})$ can be interpreted as the charge density of negatively charged anti-bosons.
By the same token,
$\rhoo_{++}(x)$ represents the positive charge density of bosons.
In the Heisenberg representation, operators for bosons and antiboson densities, therefore, are
\begin{equation} \label{13e}
  \begin{aligned}
\hat \rho_{bos}(x|t) &= \rhoo_{++}(x|t)=\sum_{kk'} \la v_k^+|v_{k'}^{+}\ra\la k|x\ra\la x|k'\ra\a^\+( \phi^+_k|t)\a( \phi^+_{k'}|t), \\
\hat \rho_{abos}(x|t) &=-\hat q_{abos}(x|t)=-\sum_{k,k'} \la v_{k'}^-|v_{k}^{-}\ra\la k'|x\ra\la x|k\ra\b^\+(\phi^-_{k}|t)\b(\phi^-_{k'}|t).
\end{aligned}
\end{equation}
Creation and annihilation operators in Eq.(\ref{13e}) satisfy the equations of motion
\begin{eqnarray} \label{14e}
\begin{cases}
 i\partial_t \a( \phi^+_{k}|t) =[ \a( \phi^+_{k}|t),\H^{KG} ]_+= \sum_{k'} h^{++}_{kk'} \a( \phi^+_{k'}|t)-\sum_{k'}  h^{+-}_{kk'}\b^\+( \phi^-_{k'}|t),\\
  i\partial_t \b^\+( \phi^-_{k}|t) =[ \b^\+( {\phi^-_{k}}|t),\H^{KG} ]_+ = -\sum_{k'}  h^{--}_{kk'}\b^\+( \phi^-_{k'}|t)+ \sum_{k'} h^{-+}_{kk'} \a( \phi^+_{k'}|t),
\end{cases}
\end{eqnarray}
which differ from Eqs.(\ref{3d})  for  fermions in that the last terms of the first and second equations
now have opposite signs. The required solution of  (\ref{14e}) is
 \begin{equation} \label{15e}
   \begin{aligned}
 \a( \phi^+_{k}|t) &= \sum_{k'}  \la \phi^+_k|\phi^+_{k'}(t)\ra \a( \phi^+_{k'})-\sum_{k'} \la \phi^+_k|\phi^-_{k'}(t)\ra\b^\+( \phi^-_{k'}),\q  \a( \phi^+_{k}|0)=\a( \phi^+_{k}), \\
 \b^\+( \phi^-_{k}|t) &= \sum_{k'}  \la \phi^-_k|\phi^+_{k'}(t)\ra \a( \phi^+_{k'})- \sum_{k'}   \la \phi^-_k|\phi^-_{k'}(t)\ra\b^\+( \phi^-_{k'}) ,\q  \b^\+( \phi^-_{k}|0)=\b^\+( \phi^-_{k}),
\end{aligned}
\end{equation}
where  $|\phi^{\mu'}_{k'}(t)\ra$ satisfies the one-particle Klein-Gordon equation
(\ref{1e}) with the initial condition $|\phi^{\mu'}_{k'}(0)\ra=|\phi^{\mu'}_{k'}\ra$ (see Appendix D).
Finally, evaluating $\rho_{bos}(x,t)=\la \Phi^\K_\ph (0) |\hat \rho_{bos}(x|t)|\Phi^\K_\ph (0)\ra$ and $\rho_{abos}(x,t)=\la \Phi^\K_\ph (0) |\hat \rho_{abos}(x|t)|\Phi^\K_\ph (0)\ra$
yieds the desired densities,
\begin{eqnarray} \label{15e1}
\rho^\ph_{bos}(x,t) =\sum_{k}\la \nm(t)|\hat \pi_+| x \ra \la x|\hat \pi_+|\nm(t)\ra+\la \ph(t)|\hat \pi_+| x \ra \la x|\hat \pi_+|\ph(t)\ra,\q\q\q\n
\rho^\ph_{abos}(x,t) =-\sum_{k} \la \pp(t)|\hat \pi_-| x \ra \la x|\hat \pi_-|\pp(t)\ra- \la \ph(t)|\hat \pi_-| x \ra \la x|\hat \pi_-|\ph(t)\ra.\q\q
\end{eqnarray}
Note that  $\rho^\ph_{abos}(x,t)$ differs from its fermionic counterpart, $\rho^\ph_{pos}(x,t)$ in Eqs.(\ref{10c1}), by the negative sign of its first term.
Different behaviour of relativistic fermions and bosons,
evident e.g. in Fig.2 of \cite{Matz},
has, therefore, been traced back to the difference between the scalar products of the corresponding one-particle theories,
$\la \phi|\psi \ra= \sum_k\la \phi|\phi_k^+\ra \la \phi_k^+|\psi\ra\pm  \sum_k\la \phi|\phi_k^-\ra \la \phi_k^-|\psi\ra$.
Next we illustrate this on a simple toy model.
\section{An exactly solvable  two-state model}
Consider the case where a two dimensional Hilbert space is spanned by two vectors,  $|\phi^+\ra$ and $|\phi^-\ra$, (the subscript $k$ is henceforth omitted).
The one-particle Hamiltonian is defined by its matrix elements,
\begin{eqnarray} \label{0x}
 \la \phi^+|\h| \phi^+\ra=E={\mp} \la \phi^-|\h| \phi^-\ra, \q  \la \phi^+|\h| \phi^-\ra=V=\la \phi^+|\h| \phi^-\ra^*.
\end{eqnarray}
and the scalar products are
\begin{eqnarray} \label{-1x}
\la \phi^+| \phi^+\ra=1, \q \la \phi^-| \phi^-\ra=\pm1,\q  \la \phi^\pm| \phi^\mp\ra =0,
\end{eqnarray}
where the upper and lower signs are for fermions and bosons, respectively.
The two cases need to be treated separately.
\begin{figure}
\includegraphics[angle=0,width=10cm]{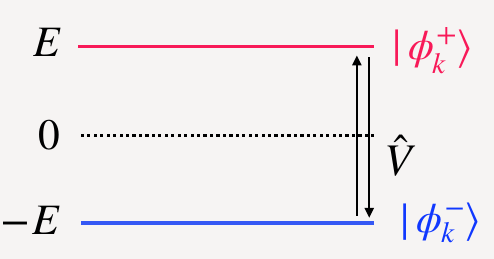}
\caption {A two-state toy model. The Hamiltonian $\h=\h_0+\hat V$ is defined by
$\h_0|\phi^\pm\ra = \pm E|\phi^\pm\ra$,
$\la \phi^\mu|\hat V|\phi^\mu\ra=0$, and $\la \phi^+|\hat V|\phi^-\ra=\la \phi^-|\hat V|\phi^+\ra^*\equiv V$.
One also has $\la \phi^\pm|\phi^\mp\ra=0$, $\la \phi^+|\phi^+\ra=1$, and $\la \phi^+|\phi^+\ra=\pm1$,
where the upper and lower signs are for fermions and bosons, respectively.
}
\label{FIG4}
\end{figure}
 \subsection{Fermions}
With no extra \e{electron} present, for the two densities in Eq.(\ref{10c1}) one finds ($\p_\mu=|\phi^\mu\ra\la \phi^\mu|$, $\mu=\pm$)
 \begin{equation} \label{1x}
   \begin{aligned}
\rho_{el}(x,t) &=\la\phi^-(t)|\hat \pi_+| x \ra \la x|\hat \pi_+|\phi^-(t)\ra =\la\phi^+|x\ra\la x|\phi^+\ra
 \la \phi^+|\phi^-(t)\ra|^2,\\
\rho_{pos}(x,t) &=\la\phi^+(t)|\hat \pi_-| x \ra \la x|\hat \pi_-|\phi^+(t)\ra =\la{\phi^-}|x\ra\la x|{\phi^-}\ra | \la \phi^-|\phi^+(t)\ra|^2.
\end{aligned}
\end{equation}
The scalar products $\la  \phi^{\pm}|\phi^{\mu}(t)\ra$ satisfy (see Appendix C)
\begin{eqnarray} \label{2x}
    i\partial_t & \begin{pmatrix}
          \la \phi^+|\phi^\mu (t)\ra  \\
            \la \phi^-|\phi^\mu(t)\ra
         \end{pmatrix}
         = &\begin{pmatrix}
         E & V\\
          V^* & -E
         \end{pmatrix}
         \begin{pmatrix}
          \la \phi^+|\phi^\mu(t)\ra  \\
            \la \phi^-|\phi^\mu(t)\ra
         \end{pmatrix}
\end{eqnarray}
For a constant $V(t)=V$ Eqs.({\ref{2x}}) can be solved  to yield
\begin{equation} \label{3x}
  \begin{aligned}
\la \phi^+|\phi^+ (t)\ra &= \om_F^{-1}[-iE\sin(\om_Ft)+{ \om_F}\cos(\om_Ft)]\\
\la \phi^-|\phi^+ (t)\ra & = -i\om_F^{-1}V\sin(\om_Ft) \\
\la \phi^+|\phi^- (t)\ra & = -i\om_F^{-1}V^*\sin(\om_Ft) \\
\la \phi^-|\phi^-(t)\ra & =\om_F^{-1}[iE\sin(\om_Ft)+{\om_F}\cos(\om_Ft)],
\end{aligned}
\end{equation}
where
\begin{eqnarray} \label{4x}
\om_F=\sqrt{E^2+|V|^2}
\end{eqnarray}
With this,  \e{electron} and \e{positron} densities and numbers become
\begin{equation} \label{5x}
  \begin{aligned}
\rho_{el}(x,t) & =\omm^{-2}|V|^2\sin^2(\omm t)\la\phi^+|x\ra\la x|\phi^+\ra,\\
\rho_{pos}(x,t) & =\omm^{-2}|V|^2\sin^2(\omm t)\la\phi^-|x\ra\la x|\phi^-\ra,\\
N_{el}(t) \equiv  \int  & \rho^{el}(x,t)dx = \omm^{-2}|V|^2\sin^2(\omm t)=N_{pos}(t).
\end{aligned}
\end{equation}
Figure \ref{FIG5} shows $N^{el}(t)$ for different values of parameters. In the case of resonance, $E=0$, the system performs complete Rabi oscillations, $N^{el}(t)=\sin^2(|V|t)=N^{pos}(t)$, with the Rabi frequency
$\omm$ equal to the absolute value of the matrix element $V$.
In general, the number of \e{electrons} an \e{positrons} cannot exceed $1$, since at $t=0$ there is a single particle,
and the number of particles is conserved.
\begin{figure}
\includegraphics[angle=0,width=8cm]{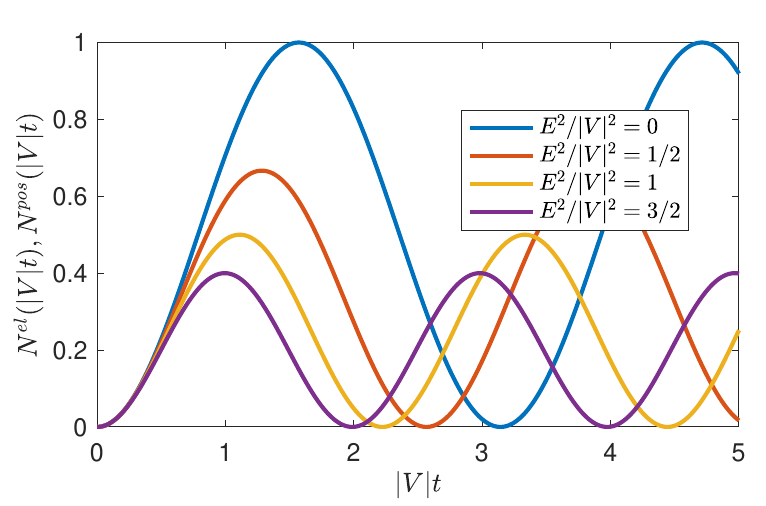}
\caption{Number of electrons (or  positrons) (\ref{5x}) for a two-level model in Fig.\ref{FIG4}, for a constant potential $V$ acting between $t=0$ and $t$,
and {$E^2/|V|^2= 0, 1/2, 1, 3/2$.}}
\label{FIG5}
\end{figure}
\newline
The case of an extra \e{electron} added at $t=0$ in the state $|\phi^+\ra$ is of no particular interest.
By symmetry, both levels remain occupied at all times, and there is always one  \e{electron}, and no \e{positron}.

 \subsection{ Bosons}
With no extra boson present, for the two densities in Eq.(\ref{10c1}) one finds ($\p_\mu=|\phi^\mu\ra\la \phi^\mu|$, $\mu=\pm$)
 \begin{equation} \label{1z}
   \begin{aligned}
\rho_{bos}(x,t) &=\la\phi^-(t)|\hat \pi_+| x \ra \la x|\hat \pi_+|\phi^-(t)\ra =\la\phi^+|x\ra\la x|\phi^+\ra | \la \phi^+|\phi^-(t)\ra|^2\\
\rho_{abos}(x,t)& =-\la\phi^+(t)|\hat \pi_-| x \ra \la x|\hat \pi_-|\phi^+(t)\ra =-\la{\phi^-}|x\ra\la x|{\phi^-}\ra | \la \phi^-|\phi^+(t)\ra|^2.
\end{aligned}
\end{equation}
Scalar products $\la  \phi^{\pm}|\phi^{\mu}(t)\ra$ satisfy (see Appendix C)
\begin{eqnarray} \label{2z}
    i\partial_t & \begin{pmatrix}
          \la \phi^+|\phi^\mu (t)\ra  \\
            \la \phi^-|\phi^\mu(t)\ra
         \end{pmatrix}
         = &\begin{pmatrix}
         E & -V\\
          V^* & -E
         \end{pmatrix}
         \begin{pmatrix}
          \la \phi^+|\phi^\mu(t)\ra  \\
            \la \phi^-|\phi^\mu(t)\ra
         \end{pmatrix}
\end{eqnarray}
For a constant $V(t)=V$, Eqs.(\ref{2z}) can be solved  to yield
\begin{equation} \label{3z}
  \begin{aligned}
\la \phi^+|\phi^+ (t)\ra &= \om_B^{-1}[-iE\sin(\om_Bt)+\om_B\cos(\om_Bt)] \\
\la \phi^-|\phi^+ (t)\ra &= -i\om_B^{-1}V\sin(\om_Bt),\\
\la \phi^+|\phi^- (t)\ra &= -i\om_B^{-1}V^*\sin(\om_Bt),\\
\la \phi^-|\phi^-(t)\ra &=-\om_B^{-1}[iE\sin(\om_Bt)+\om_B\cos(\om_Bt)]
\end{aligned}
\end{equation}
where
\begin{eqnarray} \label{4z}
\om_B=\sqrt{E^2-|V|^2}.
\end{eqnarray}
For the densities of  \e{bosons} and \e{antibosons}, produced  from vacuum $|\Phi(0)\ra =|\text{Vac}\ra$, $\a(\phi^+),\b(\phi^-)|\text{Vac}\ra=0$,  one finds [cf. Eqs.(\ref{5x})],
\begin{equation} \label{6z}
  \begin{aligned}
\rho_{bos}(x,t) & =\ommm^{-2}|V|^2\sin^2(\ommm t)\la\phi^+|x\ra\la x|\phi^+\ra,\\
\rho_{abos}(x,t) & =- \ommm^{-2}|V|^2\sin^2(\ommm t)\la\phi^-|x\ra\la x|\phi^-\ra,\\
N_{bos}(t) = \int  & \rho^{bos}(x,t)dx = \ommm^{-2}|V|^2\sin^2(\ommm t)=N_{abos}(t).
\end{aligned}
\end{equation}
Unlike in the case of fermions [cf. Eqs.(\ref{2x})] the matrix in the r.h.s. of Eqs.(\ref{2z}) is not hermitian, and one can distinguish between
\e{subcritical} regime, where $|V| < E$ and $ \ommm=| \ommm|$, and \e{supercritical} regime, where $|V| > E$ and $\ommm=i| \ommm|$.
Subcritical populations $N_{bos}(t)=N_{abos}(t)$, shown in Fig.\ref{Nbos}a{\r ,} oscillate, and may exceed unity for $E/2 < |V| <E$ .
Their supercritical counterparts in Fig.\ref{Nbos}b exhibit unlimited exponential growth, $N^{bos}(t)=N^{abos}(t)\approx |\ommm|^{-2}|V|^2\exp(2|\ommm|t)/4$ for $t\gg 1/|\ommm|$.
\begin{figure}[h]
\includegraphics[angle=0,width=14cm]{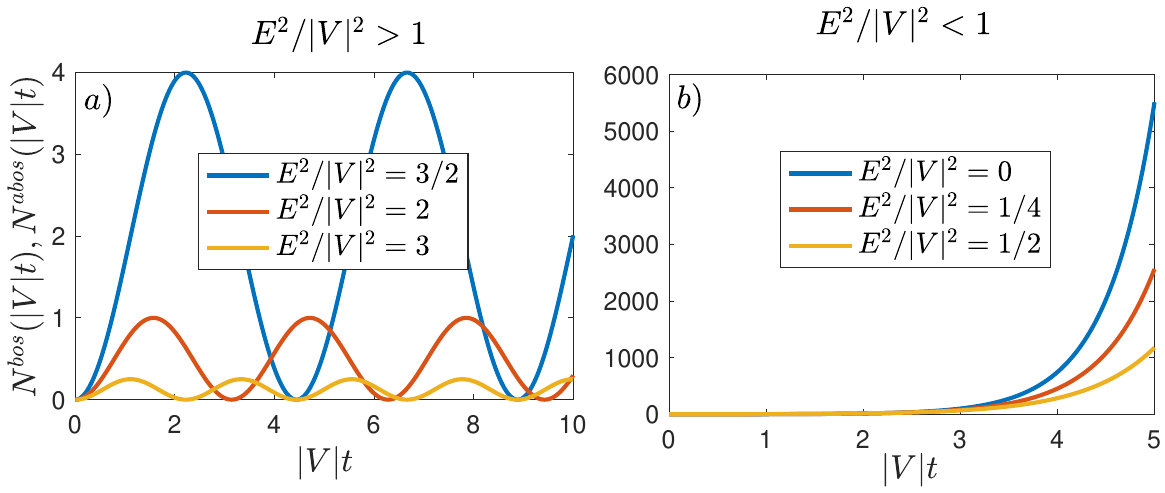}
\caption{Number of bosons and antibosons (\ref{9z}) for a two-level system in Fig.\ref{FIG4}. a) In the subcritical regime,  $|V|^2 <E^2$,
and b) in the supercritical regime $|V|^2 \ge E^2$. }
\label{Nbos}
\end{figure}
Finally, if one  boson is added at $t=0$ in the state $|\varphi\ra=|\phi^+\ra$,
$|\Phi(0)\ra =\a^\+(\phi^+)|\text{Vac}\ra$,
using Eqs.(\ref{15e1}) one obtains
\begin{equation}\label{9z}
  \begin{aligned}
N_{bos}^\varphi(t) & = \ommm^{-2}\left[ \left( E^2 + |V|^2\right) \sin^2(\ommm t) + \ommm^2 \cos^2(\ommm t)\right],\\
N_{abos}^{\varphi}(t) & = 2 \ommm^{-2}|V|^2 \sin^2(\ommm t).
\end{aligned}
\end{equation}
It is readily seen that the difference $N_{bos}^\varphi(t)-N_{abos}^{\varphi}(t)$ remains $1$ at all times.

\section{Conclusions and discussion}
We started, in a non-relativistic case, with a number of non-interacting particles.
In the chosen basis, each particle occupies one of the orthogonal states, while the rest of the states are left empty.
An external time-dependent field, switched on for a period of time, causes transitions between the states,
thus altering the initial configuration.
\newline
One can consider a projective measurement onto a subspace spanned by the initially occupied states, followed by a position measurement of the particle's position.
The subspace can be chosen to include all negative energy states (see Fig.\ref{FIG2}a). The first projective measurement eliminates the interference term,
and one obtains, separately, local densities of particles with positive and negative energies.
Treating the particle appearing in a previously unoccupied state as a \e{particle}, and the deficit of particles in a formerly occupied state
as a \e{hole}, one defines spatially resolved \e{particle} and \e{hole} densities, $\rho_{ptcle}(x,t)$ and $\rho_{hole}(x,t)$
both zero in the initial configuration,  $\rho_{ptcle}(x,0)=\rho_{hole}(x,0)=0$  [cf.  Eq.(\ref{a4})].
\newline
The just described one-particle-per-state initial configuration can be created for distinguishable particles as well as for non-relativistic fermions and bosons.
It was our purpose to establish whether $\rho_{ptcle}(x,t)$ and $\rho_{hole}(x,t)$ should be different for each particle type.
Simple calculations of Sects. III   and  IV demonstrated that in all three cases the two densities have the same form, affected neither by Pauli exclusion principle, or by
Bose-Einstein statistics.
\newline
One can hope that quantum statistical effects would become important, provided an extra \e{particle} in a superposition
in otherwise unoccupied (positive energy) states is added to the initial configuration. Indeed, with some of the target states occupied, one can expect the transitions (and, therefore,  production of holes)
to be quenched for fermions, and enhanced for bosons, since the latter are known for their tendency to \e{bunch} together (see e.g.\cite{bunch}).
Again a simple calculation shows this expectation to be wrong, as the densities for all types of particles remain identical [cf. Eqs.(\ref{10b})].
The number of \e{holes} does decrease in the presence of an extra \e{particle}, but only because the particle can be scattered and fill a \e{hole},
as would be the case regardless of the type of particles involved.
\newline
As an aside, we obtained a proof that in all cases  \e{particle} and  \e{hole} occupation numbers cannot exceed $1$ [cf. Eqs.(\ref{a5b}) and (\ref{a5c})].
Although for fermions it looks like a consequence of Pauli principle, it is in fact a property of the chosen initial configuration, and has little
to do with quantum statistics.
\newline
The analysis can be extended to Dirac electrons (at least in one spatial dimension) with minor modifications,
such as doubling the dimension of the Hilbert space in order to include an infinite number of negative energy states \cite{FV},
and filling the Fermi sea by acting with operators creating particles in the negative energy states upon the vacuum state
of the second quantised theory \cite{Greiner}.
With the ground state created in this manner, a \e{hole} can be taken to represent a positron, and casting Eqs.(\ref{10c}) into a symmetric form (\ref{10c1}) allows one
any explicit reference to the Fermi sea. Still it is difficult to see Eqs.(\ref{10c}) and (\ref{10c1}) as a consequence of the Pauli principle.
Indeed, filling the negative states (\ref{2c}) with one boson per state should yield the same result for the densities.
[We are far from suggesting that such a bosonic state is physically meaningful, yet it is allowed by the theory, and can in principle be discussed.]
\newline
Klein-Gordon bosons, studied here also in one dimension, is a case apart.
The reason is the negative sign acquired by the scalar product of a negative energy state with itself (for Dirac electrons the sign was a plus).
Second quantisation of the Klein-Gordon theory still can be carried out,  and renaming of creation and annihilation operators for the negative
states restores the usual bosonic commutation relations.
However, the ground state containing neither bosons, nor antibosons, cannot be constructed by acting with creation
operators on the vacuum state of the second-quantised theory, as was possible in the case of Dirac electrons.
Rather, such a state must be introduced independently, as was done in Eq.(\ref{12e}). With the theory thus complete,
spatially resolved boson and antiboson densities can be expressed in terms of solutions of the one-particle
Klein-Gordon equation. Despite the departure from the standard second quantisation procedure, the resulting
expressions for the densities in Eqs.(\ref{15e1}) are remarkably similar to those obtained for the Dirac electrons in Eqs.(\ref{10c1}).
\newline
Notwithstanding this formal similarity, the physical behaviour of relativistic fermions and bosons is very different, as demonstrated by the two-level example
of Sect. VII. This can be traced back to the already mention negative scalar product of the one-particle Klein-Gordon states.
To evaluate many-body densities one needs to evaluate projections of a time-evolved one-particle state onto
the positive and negative eigenstates of the field-free Hamiltonian.
In the case of Dirac electrons these projections satisfy a system of linear equations with a Hermitian matrix, since in Dirac theory the number of particles is conserved.
For Klein-Gordon bosons, the matrix is no longer Hermitian, since the theory conserves the total charge.
As a result bosonic densities and numbers can exhibit unlimited growth, shown in Fig.\ref{6}b.
 \newline
 Finally, our failure to find significant statistical effects in the behaviour of identical particles can be blamed on the chosen level of description.
The densities, studied here,  offer only a crude description of a many-body system. They can be refined by considering instead full-counting statistics,
more likely to be affected by fermionic or bosonic nature of the particles.
\section*{Acknowledgements}
We thank Grant PID2021-126273NB-I00 funded by MCIN/AEI/ 10.13039/501100011033 and by \e{ERDF A way of making Europe}.  We acknowledge financial support from the Basque Government Grant No. IT1470-22. MP acknowledges support from the Spanish Agencia Estatal de Investigaci\'on, Grant No. PID2019-107609GB-I00.
\section*{Appendix A. Second quantisation rules}
Below we list the formulae required for our analysis (other useful expressions can be found in Ch.6 of \cite{FeynS}).
Consider a state of $N$ distinguishable particles given by the  product
\begin{eqnarray}\label{AA1}
|\Phi^{dist}\ra =\prod _{n=1}^N |\phi_n\ra_{(n)}
\end{eqnarray}
where the $n$-th particle occupies a state $|\phi_{n}\ra$. The state, symmetrised for bosons (B), or antisymmetrised for fermions (F),
with respect to interchanging a pair of particles can be written as
\begin{eqnarray}\label{AA2}
|\Phi^{B,F}\ra =\prod _{n=1}^N \a^\+(\phi_n)|\text{vac}\ra
\end{eqnarray}
where operators $\a^\+(\phi_n)$ create a particle in state $|\phi_n\ra$, their Hermitian conjugates, $\a(\phi_n)$, destroy
 it, and the vacuum state $|\vac\ra$ contains no particles of any kind, $\a(\phi)|\vac\ra =0$. The operators
 either commute, or anticommute \cite{FeynS},
 \begin{equation}\label{AA3}
   \begin{aligned}
[\a(\phi),\a^\+(\psi)]_\xi & \equiv \a(\phi)\a^\+(\psi)-\xi\a^\+(\psi)\a(\phi) =\la \phi|\psi\ra, \\
\q [\a(\phi),\a(\psi)]_\xi & =[\a^\+(\phi),\a^\+(\psi)]_\xi =0,
\end{aligned}
\end{equation}
where $\xi=\pm1$, for bosons and fermions, respectively \cite{FeynS}.
Creation operators \e{transform as kets}, i.e., from $|\xi\ra=\alpha |\phi\ra +\beta |\psi\ra$,
it follows that $a^\+(\xi)=\alpha a^\+(\phi) +\beta a^\+(\psi)$. Thus, if  two complete orthonormal bases,
$|f_j\ra$ and $|g_j\ra$, are related by a unitary transformation
\begin{eqnarray}\label{AA4}
|g_j\ra=\sum_{j'}|f_{j'}\ra\la f_{j'}|g_j\ra,
\end{eqnarray}
the corresponding transformations between operators are given by
\begin{eqnarray}\label{AA5}
\a^\+(g_j)= \sum_{j'} \la f_{j'}|g_j\ra \a^\+(f_{j'}), \q \a(g_j)= \sum_{j'} \la f_{j'}|g_j\ra^* \a(f_{j'})=\sum_{j'}{\la g_j |f_{j'}\ra } \a(f_{j'}).
\end{eqnarray}
A sum of identical one-particle operators, each acting on the variables of one of the distinguishable particles,
\begin{eqnarray}\label{AA6}
\A^{dist}=\sum_n \A_{(n)}, \q \A_{(n)} \equiv \sum_{jj'}|f_j\ra_{(n)} A_{jj'}\la f_{j'}|_{(n)},
\end{eqnarray}
is replaced by an expression in which the bras and kets are replaced by the corresponding annihilation and creation operators,
respectively,
\begin{eqnarray}\label{AA6}
\A=\sum_{jj'} A_{jj'}\a^\+(f_j)\a(f_{j'}).
\end{eqnarray}
\section*{Appendix B. Derivation of Eqs. (\ref{10b1}) and (\ref{10c1}) }
In the non-relativistic case  one has,
\begin{eqnarray}\label{AD1}
\sum_n|n(t)\ra\la n(t)|+\sum_p|p(t)\ra\la p(t)|=\1,
\end{eqnarray}
and
\begin{equation}\label{AD2}
  \begin{aligned}
 \sum_n|\la x|\p_{-}|n(t)\ra|^2 & =
 \sum_n |\la x |n\ra|^2-\sum_p|\la x|\p_{-}|p(t)\ra\la p(t)|\pi_|x\ra\\
  & =\sum_n |\la x |n\ra|^2-\sum_p|\la x|\p_{-}|p(t)\ra|^2.
\end{aligned}
\end{equation}
Inserting (\ref{AD2}) into (\ref{10b}) yields (\ref{10b1}).
\newline
In the case of Dirac electrons of Sect.V the completeness relation reads
\begin{eqnarray}\label{AD3}
\sum_k|\phi_k^+(t)\ra\la \phi_k^+(t)|+\sum_k|\phi_k^-(t)\ra\la \phi_k^-(t)|=\1.
\end{eqnarray}
Introducing in $\mathcal{H}_2$ an orthonormal basis $|\nu_i\ra$, $i=1,2$, $\sum_i|\nu_i\ra\la \nu_i|=\1_2$,
one obtains
\begin{eqnarray}\label{AD4}
\pi_x=|x\ra\la x|\sum_i|\nu_i\ra\la \nu_i|\equiv \sum_i|x,i\ra \la x,i|.
\end{eqnarray}
Then
\begin{equation}\label{AD5}
  \begin{aligned}
& \sum_k\la \phi^-_k(t) | \p_{-}|x\ra \la x|\p_{-}|\phi^-_k(t)\ra=\sum_{i,x}|\la x,i|\p_{-}|\phi^-_k(t)\ra\la \phi^-_k(t)|\pi_-|x,i\ra=\\
& \sum_k\la \phi^-_k|x\ra\la x |\phi^-_k\ra
 -\sum_{i,k}\la x,i|\p_{-}| \phi^+_k(t)\ra\la  \phi^+_k(t)|\pi_-|x,i\ra =\\
 & \sum_k\la \phi^-_k|x\ra\la x |\phi^-_k\ra- \sum_k\la \phi^+_k(t) | \p_{-}|x\ra \la x|\p_{-}|\phi^+_k(t)\ra.
\end{aligned}
\end{equation}
Inserting (\ref{AD5}) into (\ref{10c}) yields (\ref{10c1}).
{\color{black} {\section*{ Appendix C. Derivation of Eqs.(\ref{10c1}) and (\ref{15e1}) }}
Both equations can be derived by using the Heisenberg equations of motion, (\ref{5d}) and (\ref{15e}),  for the corresponding creation and annihilation operators.
\subsection{Electron/boson density}
From Eqs.(\ref{5d1}) and (\ref{13e}), the density for the electrons/bosons is
\begin{eqnarray}\label{AB1}
\rho_{el,bos}^\ph(x,t)
=\la \phi^{D,KG}_\ph(0)| \sum_{kk'}\la \phi_{k}^+|x \ra \la x|\phi_{k'}^+\ra \a^\dag(\phi_{k}^+|t)\a(\phi_{k'}^+|t)|\Phi^{D,KG}_\ph(0)\ra,
\end{eqnarray}
with
\begin{equation}
  \begin{aligned}
\a^\dag(\phi_{k}^+|t) &= \sum_{k'} \left [u_{kk'}^{++}(t)^* \a^\dag(\phi_{k'}^+) - \xi\u_{kk'}^{+-}(t)^*\b(\phi_{k'}^-)\right],\\
\a(\phi_{k'}^+|t)& = \sum_{k''} \left [u_{k'k''}^{++}(t) \a(\phi_{k''}^+) - \xi\u_{k'k''}^{+-}(t)\b^\+(\phi_{k''}^-)\right],
\end{aligned}
\end{equation}
where $u_{kk'}^{\mu \mu'}(t) \equiv \la \phi_k^\mu|\phi_{k'}^{\mu'}(t)\ra$, $\xi =\pm1$ for bosons and fermions,
and the initial state is $| \Phi_\ph^{D,KG}(0)\ra = \a^\dag(\ph)|\Vac\ra$. The only non-zero contributions, therefore, are
\begin{equation}\label{AB2}
\begin{aligned}
& \la{\Vac}|\hat{a}(\varphi) \hat{a}^\dag(\phi_{k'}^+)\hat{a}(\phi_{k''}^+)\hat{a}^\dag(\varphi)| \Vac \ra = \\
& \la \Vac | \left[ \la{\varphi|\phi_{k'}^+}\ra+\xi \hat{a}^\dag(\phi_{k''}^+)\hat{a}(\varphi)\right]
\left[ \la{\phi_{k''}^+|\varphi}\ra + \xi \hat{a}^\dag(\varphi)\hat{a}(\phi_{k''}^+)\right] | \Vac \ra
 = \braket{\varphi|\phi_{k'}^+}\braket{\phi_{k''}^+|\varphi},
 \end{aligned}
\end{equation}
and
 \begin{equation}\label{AB3}
 \begin{aligned}
 &\la \Vac |\hat{a}(\varphi) \hat{b}(\phi_{k'}^-)\hat{b}^\dag(\phi_{k''}^-)\hat{a}^\dag(\varphi)| \Vac \ra = \\
& \la \Vac |\hat{a}(\varphi) \left[ \delta_{k'k''} + \xi \hat{b}^\dag(\phi_{k''}^-)\hat{b}(\phi_{k'}^-)\right] \hat{a}^\dag(\varphi)| \Vac \ra =  \delta_{k'k''},
 \end{aligned}
\end{equation}
where we have used that $\xi^2 = 1$.

Inserting (\ref{AB2}) and (\ref{AB3}) into (\ref{AB1}) and using $\sum_k |\phi_k^+(t) \ra \la \phi_k^+|\ph\ra = |\ph(t)\ra$ yields
\begin{eqnarray}\label{AB4}
\rho_{el,bos}^\ph(x,t) = \sum_k \la \phi_k^-(t)|\p_+|x\ra \la x|\p_+|\phi_k^-(t)\ra + \la \ph(t)|\p_+|x\ra \la x|\p_+| \ph(t)\ra.
\end{eqnarray}
\subsection{Positron/antiboson density}
From Eqs.(52) and (67), the density for the positrons/antibosons is
\begin{equation}\label{AB5}
\begin{aligned}
\rho_{pos,abos}^\ph(x,t) & =  \pm \la \Phi^{D,KG}_\ph(0)| \sum_{kk'} \la \phi_{k'}^-|x\ra \la x|\phi_{k}^-\ra\hat{b}^\dag(\phi_{k}^-|t)\hat{b}(\phi_{k'}^-|t)|\Phi^{D,KG}_\ph(0)\ra,\\
\b^\dag(\phi_{k}^-|t) &= \sum_{k'} \left [u_{kk'}^{-+}(t) \a(\phi_{k'}^+) - \xi\u_{kk'}^{--}(t)\b^\+(\phi_{k'}^-)\right],\\
\b(\phi_{k'}^-|t) &= \sum_{k''} \left [u_{k'k''}^{-+}(t)^* \a^\+(\phi_{k''}^+) - \xi\u_{k'k''}^{--}(t)^*\b(\phi_{k''}^-)\right],
\end{aligned}
\end{equation}
where the $-$ sign for antibosons was introduced in Sect.VIB to make $\rho_{abos}^\ph(x,t)$ a particle density.
Acting as above, the densities are
\begin{eqnarray}\label{AB6}
\rho_{pos,abos}^\ph(x,t) = \pm \sum_k \la \phi_k^+|\p_-|x\ra\la x|\p_-|\phi_k^+(t)\ra- \la \ph(t)|\p_-|x\ra \la x|\p_-|\ph(t)\ra.
\end{eqnarray}
Combining Eqs(\ref{AB4}) and (\ref{AB6}) yields Eqs.(\ref{10c1}) and (\ref{15e1}) for Dirac electrons and Klein-Gordon bosons, respectively.
\section*{ Appendix D. One-particle states for free relativistic fermions and bosons}
For free particles, the sums over $k$ should be replaced by integrals $\sum_k \to \int_{-\infty}^{\infty} dk$.
One can introduce a coordinate basis $|x,i\ra$, $i=1,2$ spanning the entire Hilbert space $\mathcal H{_2}$,
\begin{eqnarray}\label{AE0}
\la x,1|x',1\ra=\delta(x-x'), \q \la x,1|x',2\ra =\la x,2|x',1\ra=0, \q \la x,2|x',2\ra=\pm\delta(x-x'),
\end{eqnarray}
where the upper and lower signs are for fermions and bosons respectively.
Any state $|\psi\ra =\int  [|x,1\ra \la x,1|\psi\ra \pm |x,2\ra \la x,2|\psi\ra] dx$ can be represented by a two-component wave function with
coefficients $ \la x,1|\psi\ra$ and $ \la x,2|\psi\ra$.
For free one-particle Dirac states one has \cite{Greiner}
\begin{eqnarray}\label{AE1}
\begin{pmatrix}\la{x,1|\phi_k^\pm}\ra \\\la{x,2|\phi_k^\pm}\ra \end{pmatrix} =& N_k^{F \pm}\begin{pmatrix}  1 \\ \frac{k}{\pm E_k+m} \end{pmatrix}e^{ikx},
\end{eqnarray}
where $N_k^{F \pm} = \sqrt{\frac{\pm E_k+m}{\pm 4\pi E_k}}$,
and $\la{\phi_k^\mu|\phi_{k'}^{\mu'}}\ra = \int dx \sum_{i=1}^2 \la {\phi_k^\mu|x,i}\ra\la {x,i|\Phi_{k'}^{\mu'}}\ra = \delta(k-k')\delta_{\mu\mu'}$.
\newline
Free one-particle Klein-Gordon states in the Feschbach-Villars representation are given by
\begin{eqnarray}\label{AE2}
\begin{pmatrix}\la{x,1|\phi_k^\pm}\ra \\ \la {x,2|\phi_k^\pm}\ra \end{pmatrix} =& N_k^B \begin{pmatrix}  m \pm E_k  \\ m \mp E_k
\end{pmatrix}e^{ikx},
\end{eqnarray}
where $N_k^B =\frac{1}{2\sqrt{2\pi E_k m}}$, and
$\la{\phi_k^\mu|\phi_{k'}^{\mu'}}\ra = \int dx \left[ \la{\phi_k^\mu|x,1}\ra\la {x,1|\phi_{k'}^{\mu'}}\ra - \la{\phi_k^\mu|x,2}\ra \la{x,2|\phi_{k'}^{\mu'}}\ra\right] = \lambda_\mu\delta(k-k')\delta_{\mu\mu'}$, with $\lm_{\pm}=\pm1$.
\newline
With the help of Eqs.(\ref{AE1}) and (\ref{AE2}) one can evaluate  quantities like
\begin{eqnarray}\label{AE3}
\la \phi^\mu_{k}| x \ra \la x| \phi^{\mu'}_{k'}\ra=\la  \phi^\mu_{k}|x,1\ra\la x,1|\la x| \phi^{\mu'}_{k'}\ra\pm \la  \phi^\mu_{k}|x,2\ra\la x,2|\la x| \phi^{\mu'}_{k'}\ra,
\end{eqnarray}
where, as before, upper sign is for fermions, and lower sign is for bosons.
\section*{Appendix E. One-particle evolution}

Equations (\ref{15e1}) and (\ref{10c1}) can be rewritten as
\begin{equation} \label{AC0}
\begin{aligned}
\rho^\ph_{el(bos)}(x,t) &=\sum_{k,k',k''}\la \nm(t)|\phi^+_{k'}\ra \la\phi^+_{k'}| x \ra \la x| \phi^+_{k''}\ra\la \phi^+_{k''}|\nm(t)\ra+\\
&\sum_{k',k''}\la\varphi (t)|\phi^+_{k'}\ra \la \phi^+_{k'}| x \ra \la x| \phi^+_{k''}\ra\la \phi^+_{k''}|\varphi (t)\ra,\\
\rho^\ph_{pos(abos)}(x,t)& =\pm \sum_{k,k',k''}\la \pp(t)|\phi^-_{k'}\ra \la\phi^-_{k'}| x \ra \la x| \phi^-_{k''}\ra\la \phi^-_{k''}|\pp(t)\ra-\\
&\sum_{k',k''}\la\varphi (t)|\phi^-_{k'}\ra \la \phi^-_{k'}| x \ra \la x| \phi^-_{k''}\ra\la \phi^-_{k''}|\varphi (t)\ra.
\end{aligned}
\end{equation}
Consider next the one-particle equation
\begin{eqnarray}\label{AC1}
i\partial \psi(t)\ra=\h |\psi(t)\ra=[\h_0+\hat V]|\psi(t)\ra.
\end{eqnarray}
The eigenstates of $\h_0$ satisfy $\h_0|\phi^\pm_k\ra= \pm E_k|\phi^\pm_k\ra$, and we want to evaluate
scalar products $\la \phi^\pm_k|\psi(t)\ra$.
\subsection{Dirac electrons}
The eigenstates $|\phi^\mu_k\ra$, $\mu=+,-$, are subject to the scalar product $\la \phi^\mu_k|\phi^{\mu'}_{k'}\ra =\delta_{\mu\mu'}\delta_{kk'}$,
and form a complete orthonormal set,
\begin{eqnarray}\label{AC2}
\sum_k |\phi^{+}_{k}\ra\la \phi^{+}_{k}|+\sum_k |\phi^{-}_{k}\ra\la \phi^{-}_{k}| ={\hat  \1}.
\end{eqnarray}
Multiplying Eq.(\ref{AC1}) by the unit operator (\ref{AC2}) on the left, introducing the unity between $\h$ and $ |\psi(t)\ra$, and equating the coefficients
multiplying the same  $|\phi^\mu_k\ra$ on the left and right hand sides, yields the equations for $\la \phi^\mu_k|\psi(t)\ra$,
\begin{eqnarray} \label{AC3}
    i\partial_t & \begin{pmatrix}
         \underline \psi^+(t)  \\
             \underline \psi^-(t)
         \end{pmatrix}
         = &\begin{pmatrix}
         \hat h^{++} & \hat h^{+-}\\
         \hat h^{-+} & \hat h^{--}
                  \end{pmatrix}
       \begin{pmatrix}
         \underline \psi^+(t)  \\
             \underline \psi^-(t)
         \end{pmatrix},
\end{eqnarray}
where we defined column vectors
\begin{eqnarray} \label{AC4}
   \underline \psi^+(t) =
   &
   \begin{pmatrix}
         ...\\
         \la \phi^+_2|\psi(t)\ra  \\
           \la \phi^+_1|\psi(t)\ra
         \end{pmatrix}, \q
 \underline \psi^-(t) =
   &
   \begin{pmatrix}
         \la \phi^-_1|\psi(t)\ra  \\
           \la \phi^-_2|\psi(t)\ra\\
           ...
         \end{pmatrix},
\end{eqnarray}
and $\hat h^{\mu\mu'}$, $\mu,\mu'=+,-$, are the four blocks of a Hermitian matrix $\h^{\mu\mu'}_{kk'}\equiv \la \phi^\mu_k|\h|\phi^{\mu'}_{k'}\ra$.
\subsection{Klein-Gordon bosons}
Now $|\phi^\mu_k\ra$, $\mu=+,-$, are subject to scalar product $\la \phi^+_k|\phi^+_{k'}\ra =\delta_{kk'} =-\la \phi^-_k|\phi^-_{k'}\ra$,
and the completeness relation reads
\begin{eqnarray}\label{AC5}
\sum_k |\phi^{+}_{k}\ra\la \phi^{+}_{k}|-\sum_k |\phi^{-}_{k}\ra\la \phi^{-}_{k}| =\hat \1.
\end{eqnarray}
Acting as before, we obtain the equations for $\la \phi^\mu_k|\psi(t)\ra$,
\begin{eqnarray} \label{AC6}
    i\partial_t & \begin{pmatrix}
         \underline \psi^+(t)  \\
            - \underline \psi^-(t)
         \end{pmatrix}
         = &\begin{pmatrix}
         \hat h^{++} & -\hat h^{+-}\\
         -\hat h^{-+} & \hat h^{--}
                  \end{pmatrix}
       \begin{pmatrix}
         \underline \psi^+(t)  \\
             \underline \psi^-(t)
         \end{pmatrix}.
\end{eqnarray}
Multiplying Eq.(\ref{AC6})  on the left by a matrix
$\begin{pmatrix}
         \hat I  & 0\\
         0 & -\hat I
                  \end{pmatrix}$
where $\hat I$ is a unit matrix, yields the equations
for $\la \phi^\mu_k|\psi(t)\ra$,
\begin{eqnarray} \label{AC7}
    i\partial_t & \begin{pmatrix}
         \underline \psi^+(t)  \\
             \underline \psi^-(t)
         \end{pmatrix}
         = &\begin{pmatrix}
         \hat h^{++} & -\hat h^{+-}\\
         \hat h^{-+} & -\hat h^{--}
                  \end{pmatrix}
       \begin{pmatrix}
         \underline \psi^+(t)  \\
             \underline \psi^-(t)
         \end{pmatrix}
\end{eqnarray}
where the matrix in the r.h.s. is no longer unitary, owing to the opposite signs of its off-diagonal blocks.
\newline
Solutions of Eqs.(\ref{AC3}) or Eqs.(\ref{AC7}) with initial conditions $|\psi(0)\ra= |\phi^{\mu}_{k}\ra$ and  $|\psi(0)\ra= |\varphi\ra$ provide,
together with Eqs.(\ref{AE1}) and (\ref{AE2})
 the necessary data for evaluating the densities in Eqs.(\ref{AC0}).


\begin{thebibliography}{10}
        \bibitem{Grobe1} Cheng, T., Su, Q., and Grobe, R.,
Introductory review on quantum field theory with space-time resolution,
 Contemp. Phys,,  {\bf 51}, 315 (2010).
   \bibitem{Grobe2} Wagner, R.E., Ware, M.R.,  Su, Q., and Grobe, R.,
Bosonic analog of the Klein paradox,
    {\it Phys. Rev. A,} {\bf 81}, 024101 (2010).
   \bibitem{Grobe3}  Su, Q., Li, Y.T., and Grobe, R.,
Non-Perturbative Approach to Bosonic Multi??Pair Creation in Arbitrary External Fields,
    {\it Laser Physics} {\bf 22}, 745 (2012).
   \bibitem{Semi} Shockley, W. {\it Electrons and Holes in Semiconductors.} (Princeton, NJ: Van Nostrand, 1950).
    \bibitem{Schw} Schwever, S.S., {\it An introduction to relativistic field theory}, Row, Peterson and Company, (Evanston, Illinois, 1961).
    \bibitem{FeynS} Feynman, R.P., {\it Statistical mechanics}, The Benjamin/Cummings Publishing Company, (Reading, Massachusetts, 1972).
       \bibitem{BJ}  Sokolovski, D.,
Quantum Law of Rare Events for Systems with Bosonic Symmetry,
    {\it Phys.Rev.Lett.,} {\bf 110}, 115302 (2013).
        \bibitem{Shm}  Gurvitz, S.,
Wave-function approach to Master equations for quantum transport and measurement,
    {\it Front. Phys.,} {\bf 12}, 120303 (2017).
         \bibitem{Matz}  Alkhateeb, M., and Matzkin, A.,
Space-time-resolved quantum field approach to Klein-tunneling dynamics across a finite barrier,
    {\i tPhys.Rev.A,} {\bf 106}, L060202 (2022).
 \bibitem{vN} Von Neumann, J., {\it Mathematical Foundations of Quantum
Mechanics} (Princeton University Press, Princeton, 1955), pp. 183-217, Chap.
VI.
 \bibitem{bunch} Seron, B., Novo, L., and Cerf N.J.,  {Boson bunching is not maximized by indistinguishable particles},
     {\it Nature Photonics,} {\bf 17}, 702 (2023).
  \bibitem{FV} Feshbach, H., and Villars, F.,
   {Elementary relativistic wave mechanics of spin 0 and spin 1/2 particles,}
     {\it Rev. Nod. Phys.,} {\bf 30}, 24 (1958).
   \bibitem{WP} Nitta, H.,  Kudo, T., and Minowa H.,
   {Motion of a wave packet in the Klein paradox,}
     {\it Am. J.Phys.} {\bf 67}, 966 (1999).
  \bibitem{Greiner} Greiner, W., {\it Relativist quantum mechanics.}, 3rd. ed.,  (Springer, 2000).



  \end{thebibliography}
\end{document}